\documentclass[pra,aps,superscriptaddress,nofootinbib,twocolumn]{revtex4-1}

\usepackage{amsfonts}
\usepackage{amssymb}
\usepackage{amsmath}
\usepackage{graphicx}
\usepackage[usenames,dvipsnames]{color}
\usepackage[colorlinks=true,citecolor=blue,linkcolor=magenta]{hyperref}
\usepackage{ulem}
\usepackage{lmodern}
\usepackage{amsthm}

\newcommand{\figpath}{.}

\renewcommand{\Re}{\mathrm{Re}}

\newcommand{\Tr}{\mathrm{Tr}}
\newcommand{\norm}[1]{\Vert #1 \Vert}

\newcommand{\ket}[1]{\vert{ #1 }\rangle}
\newcommand{\bra}[1]{\langle{ #1 }\vert}
\newcommand{\ketbra}[2]{\vert #1 \rangle \langle #2 \vert}

\newcommand{\mean}[1]{\langle #1 \rangle}

\newcommand{\HSket}[1]{\vert{ #1 }\rangle\!\rangle}
\newcommand{\HSbra}[1]{\langle\!\langle{ #1 }\vert}
\newcommand{\HSbraket}[2]{\langle\!\langle #1 \vert #2 \rangle\!\rangle}
\newcommand{\HSketbra}[2]{\vert #1 \rangle\!\rangle \langle\!\langle #2 \vert}

\newcommand{\st}{\,\vert\,}

\newcommand{\calL}{\mathcal{L}}
\newcommand{\calG}{\mathcal{G}}
\newcommand{\calE}{\mathcal{E}}
\newcommand{\calD}{\mathcal{D}}
\newcommand{\calZ}{\mathcal{Z}}
\newcommand{\calV}{\mathcal{V}}
\newcommand{\calN}{\mathcal{N}}
\newcommand{\calM}{\mathcal{M}}

\newcommand{\RED}[1]{{\color{red} #1}}

\begin{document}
\title{Error-mitigated deep-circuit quantum simulation of open systems: steady state\\and relaxation rate problems}

\author{Anbang Wang}
\affiliation{Graduate School of China Academy of Engineering Physics, Beijing 100193, China}

\author{Jingning Zhang}
\email{zhangjn@baqis.ac.cn}
\affiliation{Beijing Academy of Quantum Information Sciences, Beijing 100193, China}

\author{Ying Li}
\email{yli@gscaep.ac.cn}
\affiliation{Graduate School of China Academy of Engineering Physics, Beijing 100193, China}

\begin{abstract}
Deep-circuit quantum computation, like Shor's algorithm, is undermined by error accumulation, and near-future quantum techniques are far from adequate for full-fledged quantum error correction. Instead of resorting to shallow-circuit quantum algorithms, recent theoretical research suggests that digital quantum simulation (DQS) of closed quantum systems are robust against the accumulation of Trotter errors, as long as local observables are concerned. In this paper, we investigate the error-mitigation problem of open systems via DQS. First, we prove that the deviation in the steady state obtained from digital quantum simulation depends only on the error in a single Trotter step, which indicates that error accumulation may not be disastrous. By numerical simulation of the quantum circuits for the DQS of the dissipative XYZ model, we then show that the correct results can be recovered by quantum error mitigation as long as the error rate in the DQS is below a sharp threshold. We explain this threshold behavior by the existence of a dissipation-driven quantum phase transition. Finally, we propose a error-mitigation technique based on the scaling behavior in the vicinity of the critical point of a quantum phase transition. Our results expand the territory of near-future available quantum algorithms and stimulate further theoretical and experimental efforts in practical quantum applications.

\end{abstract}

\maketitle
\section{Introduction}
Quantum computation~\cite{feynman1982simulating} can solve physical problems, e.g., analyzing equilibrium and dynamical properties, of strongly-correlated quantum many-body systems more efficiently than its classical counterpart. One of the prominent computational frameworks is the so-called digital quantum simulation~\cite{lloyd1996universal}, where the unitary time evolution of local Hamiltonians is first discretized into small time steps, termed as Trotter steps, and then decomposed into a number of native gates, based on the Trotter-Suzuki formulas~\cite{trotter1959product, suzuki1976relationship, suzuki1976generalized}. The quantum computational complexity, in terms of the gate count, is estimated to be polynomial in the number of constituents particles, while the classical simulation in general consumes an exponential amount of resources. With programmable quantum devices, this algorithm can be applied to a wide range of research fields, from condensed matter physics~\cite{head-marsden2021quantum} to quantum chemistry~\cite{wecker2014gate, mcardle2020quantum}. As the quantum technologies improve, the number of qubits involved in digital quantum simulation has just increased to more than 16 in recent experiments~\cite{arute2020observation, aleiner2020accurately}.

Although powerful and flexible, digital quantum simulation always uses deep circuits, with the number of quantum gates increases proportionally to the number of qubits, the evolution time, and the desired simulation accuracy. It was thus largely maintained that this algorithm would eventually be corrupted by accumulated errors, coming from both the Trotter-Suzuki expansion and imperfect quantum devices. A recent theoretical paper~\cite{heyl2019quantum}, however, suggests that digital quantum simulation is robust against Trotterization errors, in a sense that the deviations of local observables are under control for relatively large Trotter steps. Moreover, it has been shown~\cite{heyl2019quantum, sieberer2019digital} that there is a sharp threshold behavior for the Trotter errors, which can be explained by the transition from quantum localization to many-body quantum chaos. These theoretical findings have strengthened our confidence in the prospect that digital quantum simulation will beat errors and show advantages over classical simulation with noisy intermediate-scale quantum devices~\cite{preskill2018quantum}.

Alternatively, we find that there are physical problems of open quantum systems~\cite{breuer2002theory} that are robust against noises. In other words, although using deep quantum circuits, the deviations in the final results of these deep-circuit quantum simulations still remain under control, and the corresponding ideal results can be extracted with certain error-mitigation techniques~\cite{li2017efficient, temme2017error, endo2018practical}. The classical simulation of open quantum systems is harder than that of closed quantum systems, since the storage and manipulation of density matrices generally require more computational resources~\cite{noh2020efficient}, compared with those of state vectors. As a result, the universal quantum simulation of open quantum systems~\cite{kliesch2011dissipative} covers diverse quantum algorithms that are appropriate for demonstrating the potential powerfulness of quantum computers. Here in this paper, we first prove, via perturbation theory, that the deviation in the steady state obtained via digital quantum simulation depends only on the error in a single Trotter step. Using the dissipative XYZ model as an example, we then show that there is a sharp threshold for the error rate below which the ideal steady-state properties and relaxation rates can be recovered by a specific error-mitigation technique\cite{li2017efficient}, i.e., the so-called zero-noise extrapolation\cite{temme2017error}. We explain this threshold behavior by the existence of a dissipation-driven quantum phase transition, which is supported by the mean-field theory. Finally, we propose a new error-mitigation technique based on the scaling behavior~\cite{Sachdev2011quantum, Continentino2017quantum} in the vicinity of the critical point of a quantum phase transition, and provide numerical evidence with the mean-field calculation of the dissipative XYZ model. 

\section{Open quantum system} Here we briefly recall the basic description of the dynamics of open quantum systems. Consider a quantum system weakly coupled to a Markovian environment. The quantum dynamics is described by the quantum master equation of Lindblad form,
\begin{eqnarray}
\frac{d}{dt}\hat\rho(t)={\mathcal L}_0\hat\rho(t),\label{eq:master}
\end{eqnarray}
where $\hat\rho$ is the reduced density matrix of the system, and ${\mathcal L}_0$ is the Lindbladian. The general form of the Lindbladian is
\begin{eqnarray}
{\mathcal L}_0\hat\rho&=&-i\left[\hat H,\hat\rho\right]+\sum_{k=1}^K{\mathcal D}_k\hat\rho,\label{eq:lindbladian}
\end{eqnarray}
where $\hat H\equiv\sum_{l=1}^L\hat h_l$ is the system Hamiltonian and ${\mathcal D}_k\bullet\equiv\gamma_k\left(\hat A_k\bullet\hat A_k^\dag-\frac{1}{2}\hat A_k^\dag\hat A_k\bullet-\frac{1}{2}\bullet\hat A_k^\dag\hat A_k\right)$ is the dissipator of the system, operator $\hat A_k$ with non-negative coefficients $\gamma_i$ quantifying the dissipation strength. Given an initial state $\hat\rho(t=0)\equiv\hat\rho_{\rm in}$, the reduced density matrix at time $t$ can be formally obtained as
\begin{eqnarray}
\hat\rho(t)=e^{{\mathcal L}_0t}\hat\rho_{\rm in}.
\end{eqnarray}

\section{Steady-state problem} In the theory of open quantum systems, the steady-state problem, i.e., obtaining properties of steady states of given physical models, is of particular relevance. The quantum master equation in Eq.~(\ref{eq:master}) can be formally solved to obtain the reduced density matrix of the system at time $t$ as $\hat\rho(t)=e^{{\mathcal L}_0t}\hat\rho_{\rm in}$, with $\hat \rho_{\rm in}$ being the initial density matrix. Intuitively, $\hat\rho(t)$ eventually converges to the steady state $\hat\rho_0$, which is by definition annihilated by the Lindbladian, i.e., ${\mathcal L}_0\hat\rho_0=0$. In this paper, we consider only open system with single steady state, therefore the steady state $\hat\rho_{\rm 0} = \lim_{t\rightarrow\infty} e^{{\mathcal L}_0t}\hat\rho_{\rm in}$ do not depend on the initial density matrix $\rho_{\rm in}$. To investigate the speed of the convergence, it is convenient to use the Hilbert-Schmidt space, where an arbitrary operator $\hat A$, which is a $d\times d$ matrix in the Hilbert space, is mapped to $d^2$-entry column vectors $\HSket{A}$, with $d$ being the dimension of the Hilbert space and the inner product $\HSbraket{A}{B}$ defined as ${\rm Tr}\left(A^\dag B\right)$. In this representation, the Lindbladian ${\mathcal L}_0$ in Eq.~(\ref{eq:lindbladian}) becomes a $d^2\times d^2$ matrix ${\mathbf L}_0$, and the steady state $\HSket{\rho_0}$ is the right eigenvector of ${\mathbf L}_0$ with the eigenvalue $\lambda_0=0$. Formally solving the Eq.~(\ref{eq:master}) in the Hilbert-Schmidt space, we can prove that 
\begin{eqnarray}
\left\|e^{{\mathcal L}_0t}\hat\rho_{\rm in}-\hat\rho_0\right\|={\mathcal O}\left(e^{-\Gamma t}{\rm poly}(t)\right),\label{eq:convergence}
\end{eqnarray}
with an arbitrary norm. The lowest relaxation rate $\Gamma = {\rm min}\left\{-{\rm Re}\left(\lambda_\alpha\right)|\alpha\neq 0\right\}$ determines the speed of the convergence, where $\lambda_\alpha$'s are the eigenvalues of ${\mathbf L}_0$. (see Appendix~\ref{app:jf}). 

Obtaining the steady state by a classical computer needs to manipulate ${\mathbf L}_0$ and storing $\HSket{\rho}$. The classical computation costs exponential computational resources both in space and time with respect to the number of qubits $N$ in the system, since ${\mathbf L}_0$ and $\HSket{\rho}$ are respectively a $2^{2N}\times2^{2N}$ complex matrix and a $2^{2N}$ complex vector. Here we propose that the exponential cost can be alleviated by a near-term noisy quantum processor. 

\section{Quantum algorithms} A universal quantum computer can prepare the steady state by directly evolving the quantum master equation for a long enough time $T$, such that the norm $\left\|\hat\rho\left(T\right)-\hat\rho_0\right\|<\delta$, with $\delta>0$ is the predetermined accuracy goal. In contrast to the evolution of closed quantum systems, the basic operation block $e^{{\mathcal L}_0\tau}$, with $\tau$ being a small time interval to be determined by the error tolerance, cannot be represented as a unitary operator. Theoretically, a universal quantum computer can realize arbitrary quantum operations on $N$ qubits with either $2N$ ancilla qubits~\cite{nielsen2010quantum} or a single ancilla qubit with $2N$ feed-back control cycles~\cite{shen2017quantum}. Together with Eq.~(\ref{eq:convergence}), it is evident that to prepare the steady state, a universal quantum computer would require ${\mathcal O}(N)$ qubits and ${\mathcal O}\left(\log\delta^{-1}\right)$ runtime.

One of the most promising ways that universal quantum computers simulate the time evolution of quantum systems is using the Trotter-Suzuki decomposition. Here we generalize the framework proposed in Ref.~\cite{lloyd1996universal} to simulate the dynamics of open quantum systems. Specifically, we realize the following quantum operation, also called a Trotter step,
\begin{eqnarray}
e^{\mathcal L_0\tau}\simeq\ e^{{\mathcal L}^{\rm id}_{\rm eff}\tau}&\equiv&\prod_{k=1}^Ke^{{\mathcal D}_k\tau}\prod_{l=1}^L{\mathcal U}_l(\tau)\nonumber\\
&\equiv&e^{{\mathcal G}^{\rm id}_{M}\tau}\ldots e^{{\mathcal G}^{\rm id}_2\tau}e^{{\mathcal G}^{\rm id}_1\tau}
\end{eqnarray}
on the universal quantum processer, with ${\mathcal U}_l(\tau)\equiv\hat U_l(\tau)\bullet\hat U_l^\dag(\tau)\equiv e^{-i\hat h_l\tau}\bullet e^{i\hat h_l\tau}$, where ${\mathcal G}^{\rm id}_m$'s ($m=1,\ldots, M$ with $M=K+L$) are defined as the ideal native interactions being driven to implement $e^{{\mathcal D}_k\tau}$ and ${\mathcal U}_l(\tau)$. 

For near-term noisy quantum processors, the real operations can be represented as the ideal operation slightly corrupted some noise process. Mathematically, we can express the real operation as $e^{{\mathcal G}_m\tau}$ with ${\mathcal G}_m={\mathcal G}_m^{\rm id}+r_m{\mathcal E}_m$, where $r_m>0$ is a small positive value that quantifies the strength of the noise for the $m$-th quantum operation, whose generator is formally denoted as ${\mathcal E}_m$ (see Appendix~\ref{app:em} for the treatment of the depolarizing error channel). Using the Magnus expansion, the real effective Lindbladian on the near-term noisy quantum processor is
\begin{eqnarray}
{\mathcal L}_{\rm eff}&=&{\mathcal L}_0+r\sum_{m=1}^M{\mathcal E}_m+\frac{\tau}{2}\sum_{m>m'}\left[{\mathcal G}_m,{\mathcal G}_{m'}\right]\nonumber\\
&&+{\mathcal O}\left(r^2,r\tau,\tau^2\right),
\end{eqnarray}
where by definition ${\mathcal L}_0=\sum_{m=1}^M{\mathcal G}^{\rm id}_m$. Note that here we treat both $r$ and $\tau$ as small quantities of the same scale. Intuitively, the deviation of ${\mathcal L}_{\rm eff}$ from ${\mathcal L}_0$ is determined by both the noise strength $r$ and the length of the Trotter step $\tau$. Note that the high-order terms can be found in Appendix~\ref{app:me}. The effect of Trotter step $\tau$ has been widely studied and it has been shown that there is a sharp threshold for $\tau$ for closed system~\cite{heyl2019quantum}. There is also such a threshold for $\tau$ in open systems, however, we choose a fixed $\tau=0.01$ below the threshold and focus on the unexploited field on the effect of noise strength $r$.

When the total evolution time $T$ is sufficiently long, the state of the system converges to the steady state of the real effective Lindbladian, which satisfies ${\mathcal L}_{\rm eff}\hat\rho_{\rm eff}=0$. Using Eq.~(\ref{eq:convergence}) and the triangle inequality, we find the difference between the final state on the noisy quantum processor and the ideal steady state satisfies the following inequality,
\begin{eqnarray}
\left\|e^{{\mathcal L}_{\rm eff}T}\hat\rho_{\rm in}-\hat\rho_0\right\|&\leq&\left\|\hat\rho_{\rm eff}-\hat\rho_0\right\|\nonumber\\
&&+{\mathcal O}\left(e^{-\Gamma_{\rm eff}T}{\rm poly}(T)\right),
\end{eqnarray}
where $\Gamma_{\rm eff}$ is the lowest relaxation rate of the real effective Lindbladian ${\mathcal L}_{\rm eff}$. The second term in the above equation can be arbitrarily suppressed by increasing the evolution time $T$, thus the error in the prepared steady state is dominated by the difference between $\hat\rho_{\rm eff}$ and $\hat\rho_0$. Next, we will show that the steady state $\hat\rho_{\rm eff}$ and the relaxation rates $\lambda_{{\rm eff},\alpha}$ of the noisy effective Lindbladian ${\mathcal L}_{\rm eff}$ can be obtained from the corresponding ideal ones by perturbation theory, thus the deviations of the prepared steady state and the measured relaxation rates from the ideal ones are bounded by the error of a single Trotter step, even though the quantum algorithms consist deep quantum circuits.

\section{Zero-noise extrapolation} To use the perturbation theory, we first write ${\mathcal L}_{\rm eff}={\mathcal L}_0+{\mathcal L}'$ and treat ${\mathcal L}'$ as the perturbation. The steady state of $\hat\rho_{\rm eff}$ can be expressed as a power series of superoperators working on the ideal steady state $\hat\rho_0$,
\begin{equation}
\hat\rho_{\rm eff}=\sum_{n=0}^\infty(-1)^n\left({\mathcal L_0}^{-1}{\mathcal L'}\right)^n\hat\rho_0,
\end{equation}
where ${\mathcal L}_0^{-1}$ is the superoperator corresponding to the generalized inverse of ${\mathbf L}_0$ in the Hilbert-Schmidt space. Therefore, the expectation of any  observable $\hat O$ can be written as
\begin{eqnarray}
\mean{\hat O} &=& \Tr[\hat O \hat\rho_{\rm eff}] = \sum_{n=0}^\infty(-1)^n\left[\hat O\left({\mathcal L_0}^{-1}{\mathcal L'}\right)^n\hat\rho_0 \right].
\end{eqnarray}
Similarly, the relaxation rates, i.e., the real parts of the eigenvalues of the Lindbladian, can also be expressed as $\lambda_{{\rm eff}, \alpha}=\sum_{n=0}^\infty\lambda_\alpha^{(n)}$ (see Appendix~\ref{app:pt} for the expressions of $\lambda_\alpha^{(n)}$).

The power series of $\hat\rho_{\rm eff}$ will converge within a few orders under the condition $\left\|{\mathcal L}_0^{-1}{\mathcal L}'\right\|\ll 1$, with $\left\|{\mathcal L}_0^{-1}\right\|\sim 1/\Gamma$ and $\left\|{\mathcal L}'\right\|\sim r$. Thus it is anticipated that if the noise strength $r$ is smaller than the lowest relaxation rate $\Gamma$, the steady state $\hat\rho_0$ of the ideal target Lindbladian ${\mathcal L}_0$ can be inferred by extrapolation with linear or low-order polynomial functions, given noisy data obtained with the intrinsic error rate $r_0$ and boosted error rates $cr_0$ with $c>1$. (See Appendix~\ref{app:nr} for details.) In the following, we will give a concrete example of using the error mitigation technique to obtain the physical properties of the ideal steady state, with a programmable universal quantum processor.

\section{Dissipative XYZ model} We consider the dissipative version of the anisotropic spin-$\frac{1}{2}$ Heisenberg model~\cite{lee2013unconventional}, also called the XYZ model, on a two-dimensional square lattice as an example. The ideal Lindbladian ${\mathcal L}_0$ takes the general form in Eq.~(\ref{eq:lindbladian}), with the Hamiltonian for the spin lattice being
\begin{eqnarray}
\hat H_{\rm dXYZ}=\hat H_{\rm c}+\hat H_{\rm r}
\end{eqnarray}
with the row and column parts defined as,
\begin{eqnarray}
\hat H_{\rm c}=\sum_{i,j,\alpha} J_\alpha\hat\sigma_{i, j}^\alpha\hat\sigma_{i+1,j}^\alpha,\quad \hat H_{\rm r}=\sum_{i,j,\alpha} J_\alpha\hat\sigma_{i, j}^\alpha\hat\sigma_{i,j+1}^\alpha,
\end{eqnarray}
with $\hat\sigma_{i,j}^\alpha$ ($\alpha = x,y,z$) being Pauli matrices on the qubit at the $i$-th row and the $j$-th column. Here the anisotropic Heisenberg interaction strength on the nearest-neighbor sites are quantified by $J_\alpha$ ($\alpha = x,y,z$). Besides the above Hamiltonian, each spin is also subject to a dissipation process governed by the following dissipator,
\begin{eqnarray}
{\mathcal D}_{i,j}=\gamma\left(\hat\sigma_{i,j}^-\bullet\hat\sigma_{i,j}^+-\frac{1}{2}\left\{\hat\sigma_{i,j}^+\hat\sigma_{i,j}^-,\bullet\right\}\right),
\end{eqnarray}
where $\gamma$ is the intrinsic single-site dissipation rate. The steady state $\hat \rho_0$ can be obtained from an arbitrary initial state $\rho_{\rm in}$ after the following Trotterized evolution:
\begin{eqnarray}
\rho_0 = \lim_{T\rightarrow\infty}\left[\prod_m e^{{\mathcal G}_m\tau}\right]^{T/\tau}\hat\rho_{\rm in}, \label{eq:dxyzT}
\end{eqnarray}
where quantum gate $\mathcal G_m$ is either generator of local column (row) Hamiltonian $-i[J_\alpha\hat\sigma_{i, j}^\alpha\hat\sigma_{i+1,j}^\alpha, \bullet]$ ($-i[J_\alpha\hat\sigma_{i, j}^\alpha\hat\sigma_{i,j+1}^\alpha, \bullet]$),  or generator of dissipator $\mathcal D_{i,j}$.

This model plays an important role in the field of unconventional magnetism and nonequilibrium phase transition~\cite{lee2013unconventional}. In contrast to the equilibrium case, each spin in the system keeps precessing around the effective magnetic field stemming from its surrounding spins. Thus the phase transition can be understood from the aspect of the competition between the precession and the dissipation, which is quantified by the dimensionless transversal aspect ratio
\begin{eqnarray}
g\equiv\left|J_x-J_y\right|/(2\gamma). 
\end{eqnarray}
Specifically, in the thermodynamic limit, the critical point is $g_{\rm cri}=0.062$ for the dissipative XYZ model with $J_z=\gamma$~\cite{jin2016cluster,rota2018dynamical}, and the steady state lies in the paramagnetic (ferromagnetic) phase when $g<g_{\rm cri}$ ($g>g_{\rm cri}$).

\section{Experiment protocol} There are many theoretical efforts~\cite{Bacon2001universal, Sweke2014simulation, Sweke2015universal, Sweke2016digital} been made to simulate Moarkovian open quantum systems with unitary operations. Recently, experimental simulation of a two-level open quantum system is demonstrated in a superconducting circuit\cite{Han2021experimental}. It is not hard to generalize these ideas to the dissipative XYZ model.
The steady state of the dissipative XYZ model on a $L\times L$ square lattice can be prepared with a programmable quantum processor consisting of $L\times 2L$ qubits. The quantum circuit for the preparation of the steady state on a $2\times2$ spin lattice is shown in Fig.~1 as an example, where the construction only requires local interactions and can be straightforwardly generalized to $L>2$. In the Trotterized evolution \eqref{eq:dxyzT}, the order of quantum gate $\mathcal G_m$ can be chosen arbitrarily. In numerical calculation, we choose to group gates corresponding to column (row) Hamiltonian together as $e^{\mathcal G_{c(r)}\tau}$s and gates corresponding to dissipators together as $e^{\mathcal G_{d}\tau}$s as shown in Fig.~1.

\section{Small-angle gates and standardized circuits}
Having designed the circuit architecture, we can implement each module in the circuit by small-angle gates or a standard universal gate set. Mainstream physical platforms for quantum information processing, like the superconducting circuit system and the trapped ion system, have their own native gates with continuously tunable parameters. Here we take the controlled-phase (CP) gate, defined as ${\rm CP}(\theta)=\exp\left(-i\theta\ket{11}\bra{11}\right)$, in the superconducting circuit system as a representative example. With simple algebra, it is clear that ${\rm CP}(\theta)$ is equivalent to $\exp\left(-\frac{i\theta}{4}\hat\sigma_1^z\hat\sigma_2^z\right)$ and single-qubit rotations. With single-qubit $\pi/2$-rotations on the transverse axis and the CP gate, we can realize Ising-type interaction alone all three axis. (See Appendix~\ref{app:c} for details and realization of other gates.) 

\begin{figure}
\includegraphics[width=1.0\linewidth]{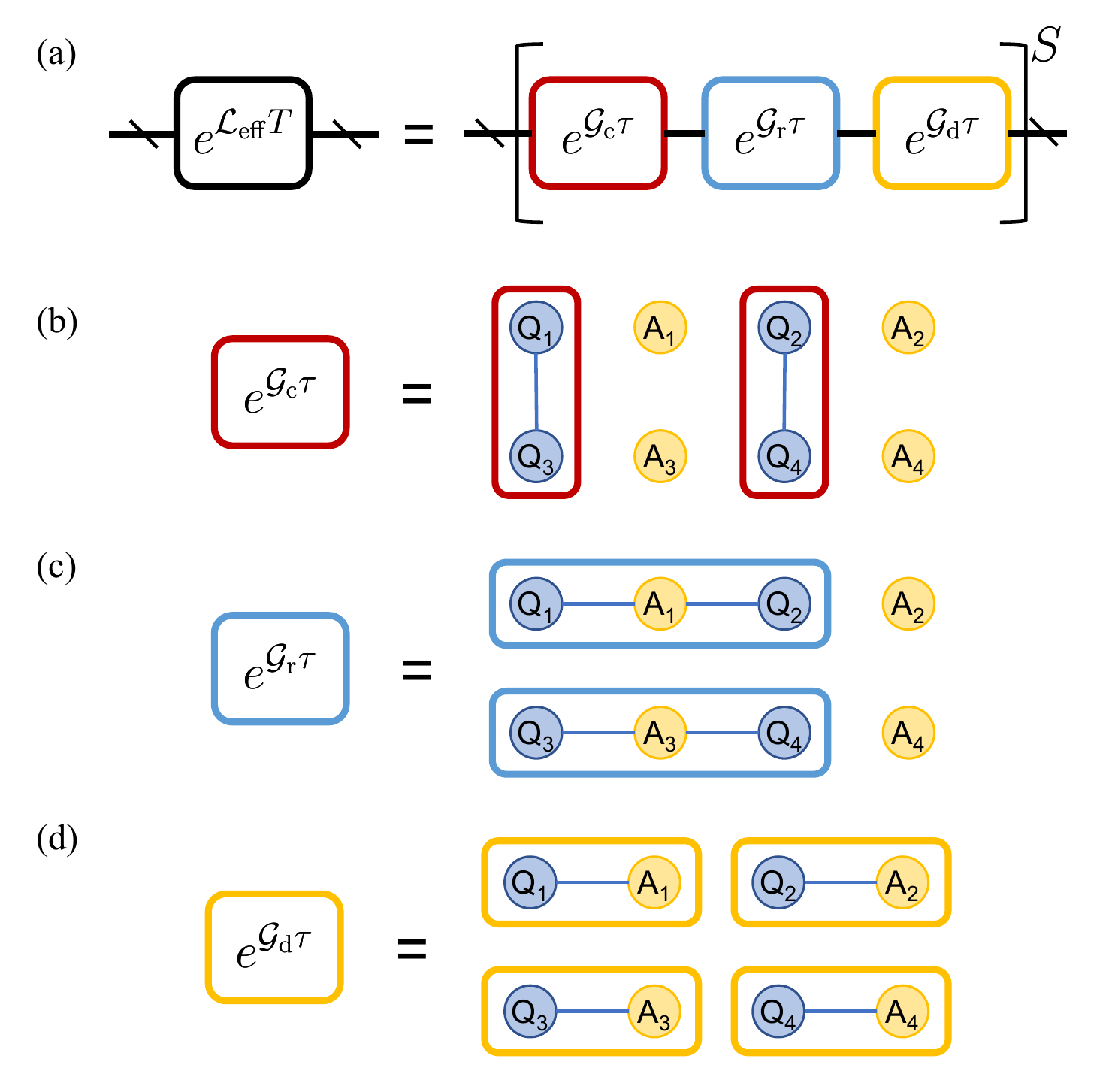}
\caption{Schematic diagram of the quantum circuit preparing the steady state. (a) Trotterized quantum circuit for the dissipative XYZ model on a 2D square lattice and evolution governed by the column Hamiltonian. (b) Qubit-ancilla layout of a $2\times2$ square lattice. There is a column of ancillas besides each column of qubits. In this layout, the column Hamiltonian $\hat H_{\rm c}$ can be directly implemented by a universal gate set among qubits. (c) Evolution governed by the row Hamiltonian. Since neighboring qubits are interleaved by ancillas in this layout, the terms in the row Hamiltonian $\hat H_r$ need to be implemented by using the ancillas as mediators. This can be achieved either by swap-like gates on the ancilla-qubit pairs or using the ancillas as tunable couplers~\cite{PhysRevLett.127.060505}. (d) Evolution governed by on-site dissipators. A recent experiment~\cite{cai2020arbitrary} shows that with a single ancilla, it is possible to implement arbitrary quantum operations on a qubit. }
\label{fig:qcircuit}
\end{figure}

\section{Noise model} We consider the depolarizing error model for the noisy quantum processor. Specifically, in the numerical simulation of the quantum circuit for the preparation of the steady states of the dissipative XYZ model, every quantum gate is followed by a quantum operation representing the depolarizing error, $e^{{\mathcal G}_m\tau}=e^{r{\mathcal E}_m\tau}e^{{\mathcal G}_m^{\rm id}\tau}$, with the following error generator
\begin{eqnarray}
{\mathcal E}_m = \left[\frac{\openone_{{\mathcal A}_m}}{2^{\left|{\mathcal A}_m\right|}}{\rm Tr}_{{\mathcal A}_m}\left(\bullet\right)-\bullet\right],
\end{eqnarray}
where ${\openone}_{{\mathcal A}_m}$ is the identity operator on the Hilbert space of the qubit set ${\mathcal A}_m$, which are involved in the $m$-th quantum operation. Note that the number of qubits in the set $\left|{\mathcal A}_m\right|=1$ (2) if the $m$-th quantum operation involves one (two) qubit(s). For simplicity, we assume a homogeneous strength for all quantum operations, i.e., $r_m=r$. We believe that introducing different noise strengths for each quantum operation would not cause quantitative changes in our conclusion.

Before presenting the numerical results, we wouldd like to make some more comments on the choice of choosing depolarizing as the representative noise model. Although being idealized, the depolarizing error model is widely used in the quantitative analysis of the performance of quantum algorithms. The reason for this choice is twofold. First, the depolarizing error model introduces all possible Pauli errors, thus provides an unbiased criterion for the algorithmic robustness. Second, by Pauli twirling~\cite{dur2005standard}, it is possible to convert generic noise into the Pauli channel, of which the depolarizing model is a special case. Finally, we mention that though we choose depolarizing error as the representative, our conclusions which will be shown below hold for various noise models. In Appendices~\ref{app:em} and~\ref{app:nr}, we present other noise models and corresponding numerical results, from which we see that the details of error mitigation for different noise models depend on their symmetries.

\begin{figure}
\includegraphics[width=1.0\linewidth]{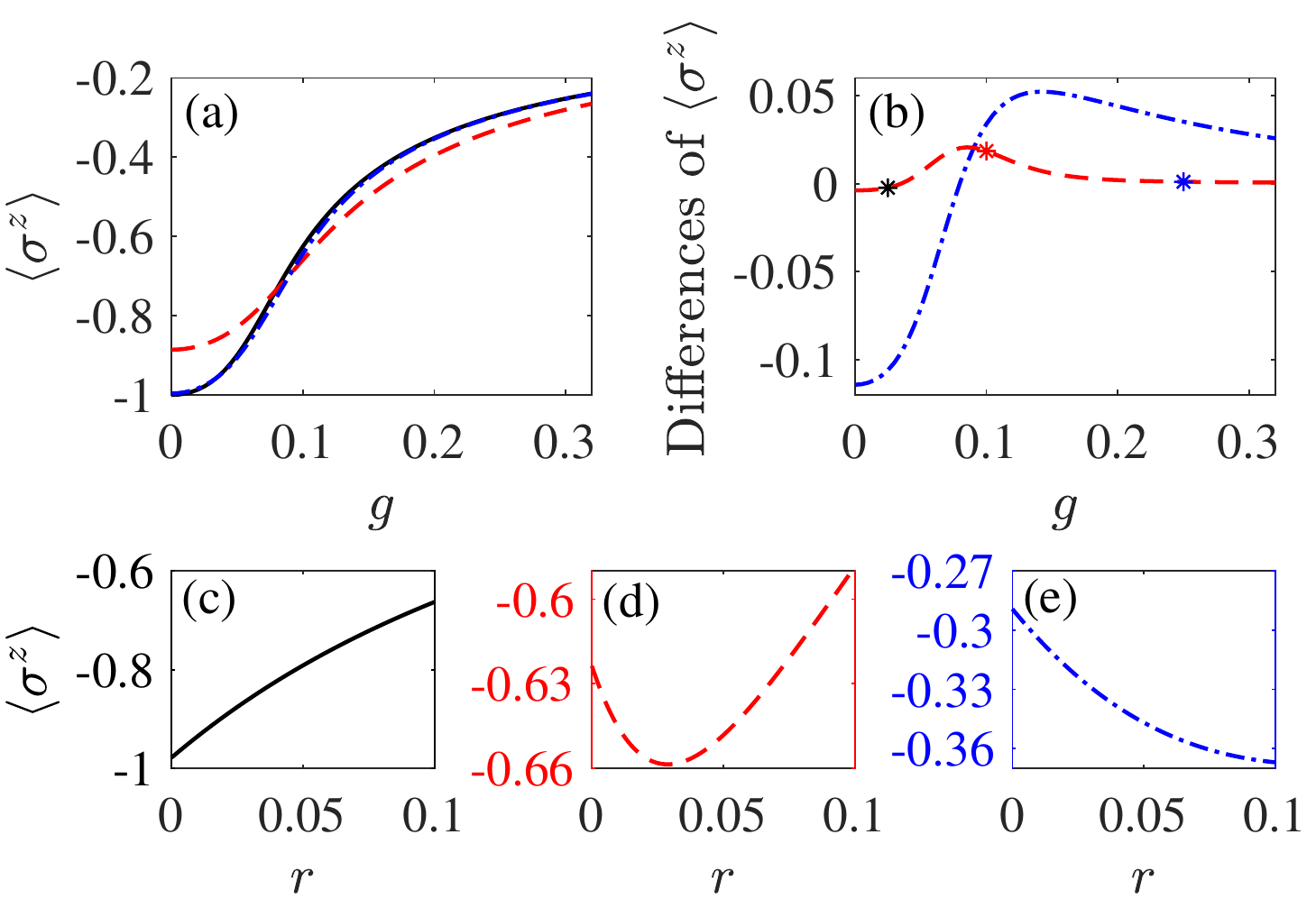}
\caption{Magnetization of a $3\times 3$ qubit lattice. (a) Curves of magnetization $M$ vs transversal aspect ratio $g$. The exact values $M_{0}$, noisy values $M_{\rm eff}$ at $r_0=0.01$, and values obtained by zero-noise extrapolation $M_{\rm ex}$ are expressed by the black solid line, blue dot-dashed line, and red dashed line. (b) Discrepancies of noisy (blue dot-dashed line) and the extrapolated (red dashed line) values of magnetization in (a). Three representative values of $g=0.025$, $0.1$ and $0.25$ are marked by the black, red, and blue asterisks. [(c)-(e)] Curves of magnetization $M$ vs noise rate $r$ at $g=0.025$, $0.1$, and $0.25$.}
\end{figure}

\section{Numerical results} We numerically simulate the process of preparing the steady state and obtaining the values of observables of the dissipative XYZ model by a noisy quantum processor. In Fig.~(2), we show the performance of the zero-noise extrapolation technique with the measured steady-state magnetization,
\begin{eqnarray}
M&\RED{\equiv}&\langle\sigma^z\rangle=\frac{1}{L^2}\sum_{i=1}^{L^2}{\rm Tr}\left[\sigma_i^z\hat\rho\right],
\end{eqnarray}
on a $3\times3$ qubit lattice. Due to the finite-size effect, the magnetization curve smoothly increases, instead of abruptly jumping, from $-1$ towards $0$, as the aspect ratio $g$ changes. Specifically, we show in the Fig.~2(a) the steady-state magnetization for the exact and the noisy prepared steady states $\hat\rho_0$ and $\hat\rho_{\rm eff}$, and the steady-state magnetization obtained by zero-noise extrapolation at $r_0 = 0.01$. To make a closer investigation, we show the discrepancies of the noisy and the extrapolated data from the true values in Fig.~2(b). We find that the discrepancy between $M_0$ and $M_{\rm eff}$ can be greatly decreased by the zero-noise extrapolation results $M_{\rm ex}$. The zero-noise extrapolation technique works quite well in the regimes far from the critical point, where the discrepancies are substantially suppressed. While in the vicinity of the critical point, the discrepancies become larger after extrapolation. To find the reason for the failure of the extrapolation, we choose three parameters, specified by stars in Fig. 2(b), and plot the magnetization as a function of the modular error rate $r$ in Figs. 2(c)-(e), respectively.
For parameters that lie deep in the paramagnetic and ferromagnetic regime, the steady-state magnetization exhibits monotonic behavior as the modular error rate $r$ increases. However, in the vicinity of the phase transition, the curve first goes down and then bends upward, which results in the failure of the extrapolation when the raw modular error rate is close to or larger than the inflection point.
The inflection point occurs at $r=0.03$ in Fig.~2(d), therefore in order for the zero-noise extrapolation technique to work, we need to choose an intrinsic error rate $r_0$ such that $2r_0<0.03$. The threshold below which the correct result of observables can be recovered by the quantum error mitigation is determined the minimum $r$ at which the inflection point occurs as $g$ varies. We find that an acceptable threshold is $r_0=0.01$. 

\begin{figure}
\includegraphics[width=1.0\linewidth]{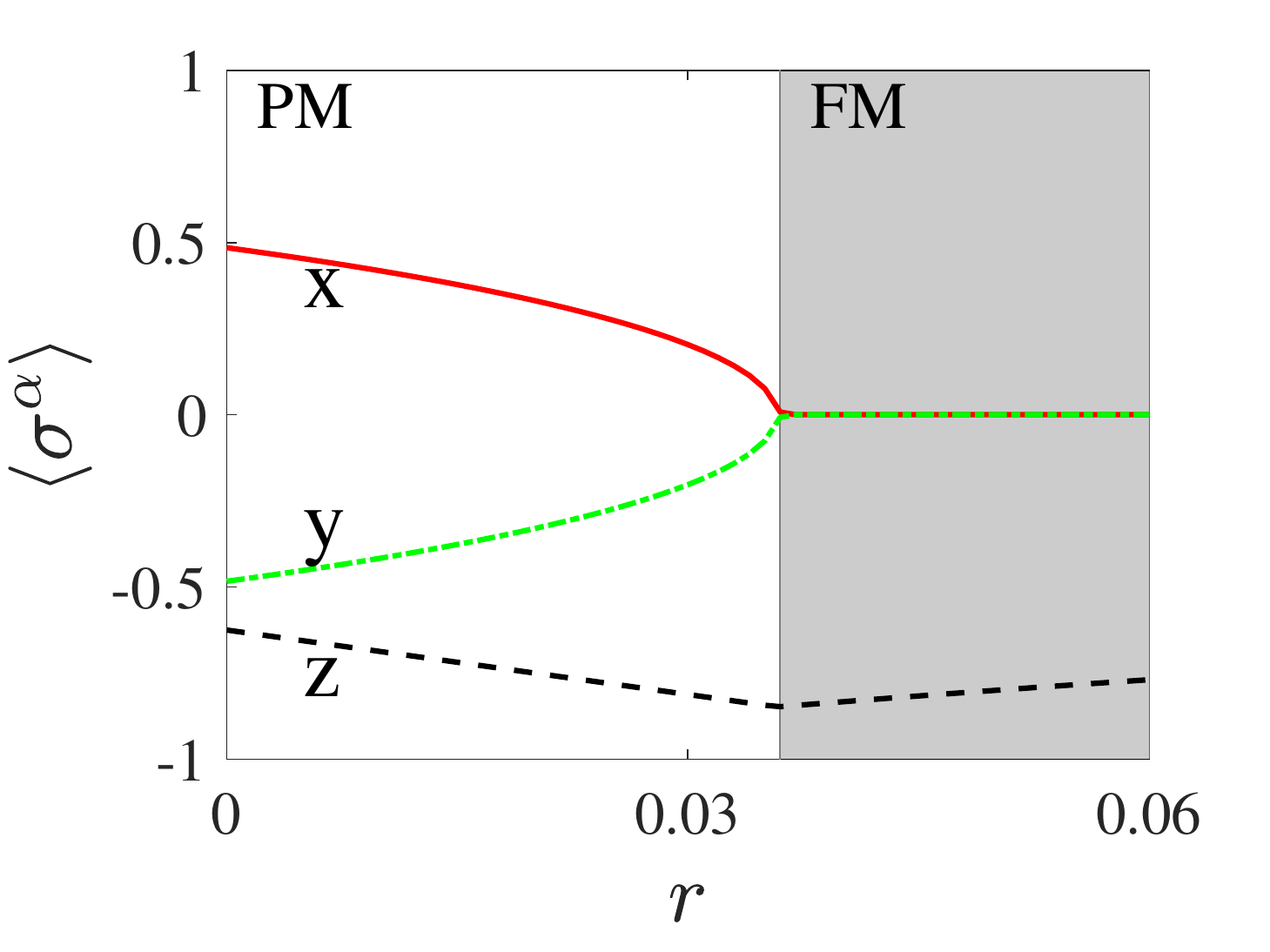}
\caption{Phase transition driven by the noise at $g=0.1$ (mean-field results). As the modular error rate $r$ increases, the system runs from paramagnetic (PM) phase to ferromagnetic (FM) phase, with order parameter $\langle\sigma^x\rangle$ and $\langle\sigma^y\rangle$ going from finite values to zeros. $\langle\sigma^z\rangle$ also goes through an abrupt change.}
\end{figure}

Then we use the mean-field method~\cite{lee2013unconventional} to reveal the physical insights behind the distinct behavior in different parameter regimes. Based on the effective Lindblad quantum master equation, the equations of motion for the expectation values of the local spin components can be written as follows,
\begin{eqnarray}
\frac{d}{dt}\mean{\hat\sigma_i^\alpha}={\rm Tr}\left[\hat\sigma_i^\alpha{\mathcal L}_{\rm eff}\hat\rho\right],
\end{eqnarray}
where $\mean{\hat\sigma_i^x}$ and $\mean{\hat\sigma_i^y}$ serve as order parameters characterizing the transition between the paramagnetic and ferromagnetic phases. We then introduce the lowest-order of the mean-field approximation to express the correlation function $\mean{\hat\sigma_i^\alpha\hat\sigma_j^\beta}$ as product of expectations $\mean{\hat\sigma_i^\alpha}\mean{\hat\sigma_j^\beta}$, in which way we obtain a set of closed nonlinear differential equations for $\mean{\hat\sigma_i^\alpha}$.  With the assumption all spins are the same, i.e., $\mean{\hat\sigma_i^\alpha}=\mean{\hat\sigma^\alpha}$, we solving the mean-field equations of motion for $\mean{\hat\sigma^\alpha}$ for different values of the modular error rate $r$, and present the numerical results in Fig. 3. Starting from the vicinity of the critical regime, with $g = 0.1$ as marked by the red asterisk in Fig. 2(b), it is clear that there is a noise-driven phase transition as $r$ increases. Moreover, the critical point $r_{\rm cri}$ predicted by the mean-field calculation is quantitatively consistent with the inflection point in Fig.~2(d). 

Thus we believe the noise-driven phase transition well accounts for the failure of the zero-noise extrapolation. Also inspired by this finding, we propose to use the scaling behavior to mitigate errors. The order parameter of the dissipative XYZ model is 
\begin{eqnarray}
m&\RED{\equiv}&\langle\sigma^x\rangle=\frac{1}{L^2}\sum_{i=1}^{L^2}{\rm Tr}\left[\sigma_i^x\hat\rho\right].
\end{eqnarray}
In the vicinity of the critical point $g_{\rm cri}$, the order parameter $m(g)$ can be described by a power-law function, $m\left(g\right)\propto \left|g-g_{\rm cri}\right|^{\beta}$, with $\beta$ being the critical exponent. Consider the case that the noises introduced by the realistic quantum processor possess the same symmetry as the model being simulated, for example, the dissipative XYZ model and the depolarizing error channel which both have the ${\mathbb Z}_2$ symmetry. (See Appendix~\ref{app:em} for other noise examples.) Then the error-perturbed order parameter will still retain the power-law dependency on not only the system parameter $g$ but also the error rate $r$, i.e.,
\begin{eqnarray}
m\left(g, r\right)\propto \left|g-g_{\rm cri}(r)\right|^{\beta(r)},\label{eq:scal}
\end{eqnarray}
where the $r$-dependency comes from the perturbation to the critical point and the critical exponent. To mitigate the effect of the errors, we can repeat the measurement of the order parameter under several different values of the noise strength $r$, e.g., the intrinsic noise strength $r=r_0$ and boosted ones $r=r' \equiv cr_0$ with $c>1$, and fit the data with power-law functions to obtain $g_{\rm cri}(r)$ and $\beta(r)$, based on which we can infer the true values $g_{\rm cri}$ and $\beta$ by extrapolation to the zero-noise limit ($r=0$).

\begin{figure}
\includegraphics[width=1.0\linewidth]{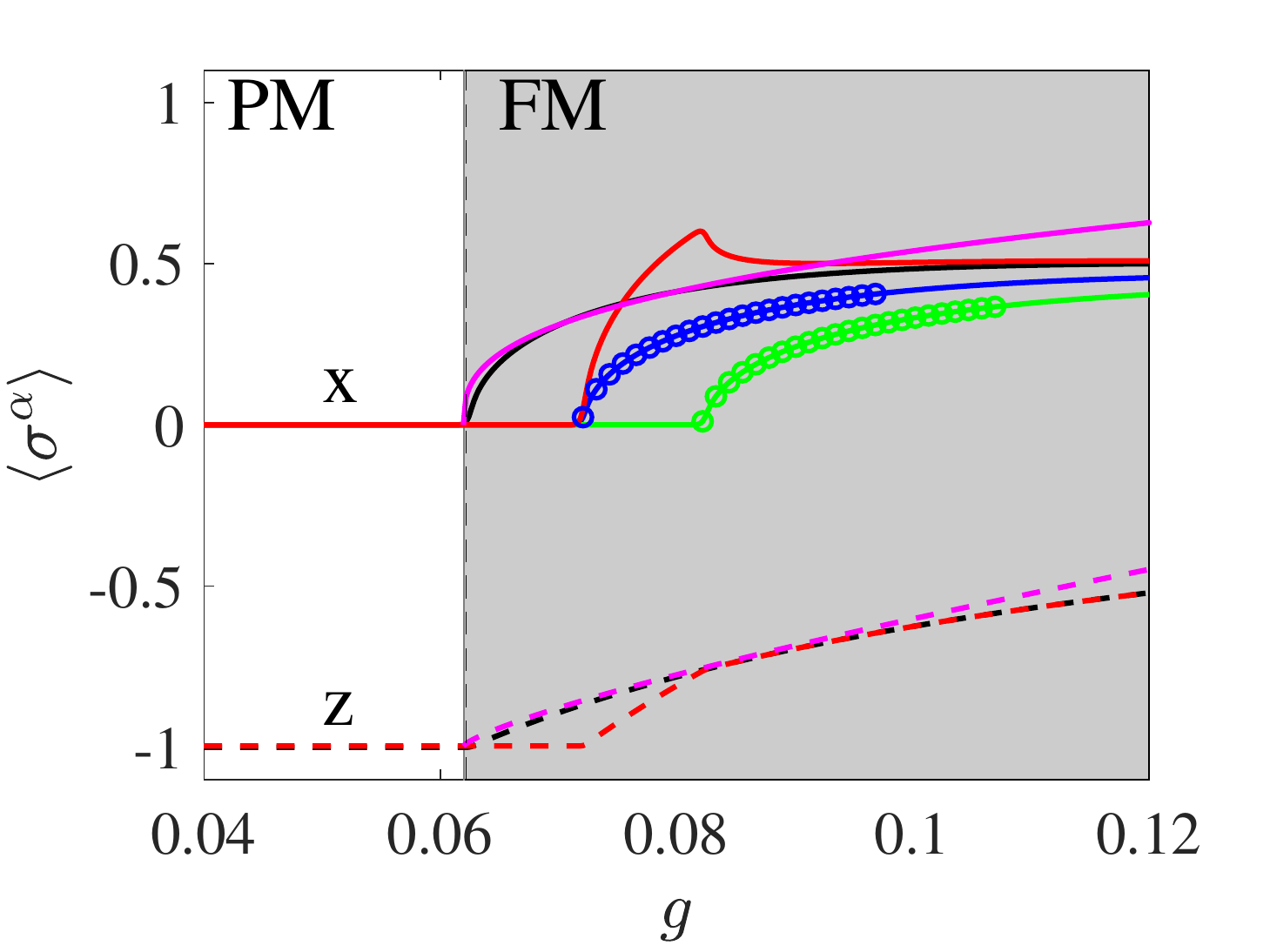}
\caption{Zero-noise extrapolation and scaling extrapolation (mean-field results)). Exact and noisy values at $r=r_0=0.01$ and $r=2r_0=0.02$ are shown by the black, blue, and green lines. The vertical dashed line is the critical point $g=0.062$ between the paramagnetic (PM) phase and ferromagnetic (FM) phase. The values obtained by zero-noise extrapolation are shown by the red line, which works well in regime away from the critical point. The circles mark the values from which we evaluate the coefficients in the power-law function of phase transition. The cyan line is the analytical curve with coefficients given by scaling extrapolation which works well near the critical point. Data of $\langle\sigma^x\rangle$ are given by the solid line and $\langle\sigma^z\rangle$ are given by the dashed line. We omit the noisy curves for $\langle\sigma^z\rangle$.}
\end{figure}

After establishing the error-mitigation technique based on the scaling behavior, we continue to investigate its performance in predicting true values of physical quantities in the whole parameter regime. We first accumulate data for the order parameters, i.e., the expectation values of the local spin components $\mean{\hat\sigma^{x}}$ and $\mean{\hat\sigma^{z}}$, for the intrinsic noise strength $r_0$ and a boosted noise strength $r'=2r_0$ using the above mentioned mean-field method. By utilizing both error-mitigation techniques, i.e., the direct extrapolation and the extrapolation based on the scaling behavior, we obtain the zero-noise results as shown in Fig.~4, and find the scaling-behavior-based extrapolation correctly recovers the true position of the critical point, while the direct extrapolation performs well on predicting the exact values of the order parameter when the parameter $g$ goes deeply into the ferromagnetic phase. (It also predicts values correctly in paramagnetic phase where the order parameter vanishes.) For comparison, we also put the true values of the order parameters. We can clearly see that the effect of the symmetry-preserving noise on the order parameter is that it moves the critical point towards the ferromagnetic phase. The failure of the scaling-behavior-based extrapolation outside the critical regime is straightforwardly connected with the invalidation of the scaling behavior outside the critical regime. 

The scaling-behavior-based extrapolation is helpful as it can be used to estimate the critical point which is of vital importance in many physical problems. We show in Fig.~5 the critical points $g_{\rm cri}$ of the dissipative XYZ model under the influence of various noises and evaluated via different extrapolations. (See Appendix~\ref{app:nr} for details of noises and extrapolations.) The estimation of the critical point predicted by the linear extrapolation is quite accurate when $r_0$ is below the threshold $0.01$. Note the deviation of the estimated critical point can be made smaller above the threshold if we use quadratic extrapolation. It seems that the scaling-behavior-based extrapolation performs well on estimating the critical point even though the value of the order parameter itself is not so well predicted when the intrinsic error rate $r_0$ exceeds the threshold. 
\begin{figure}[tbp]
\centering
\includegraphics[width=1.0\linewidth]{\figpath 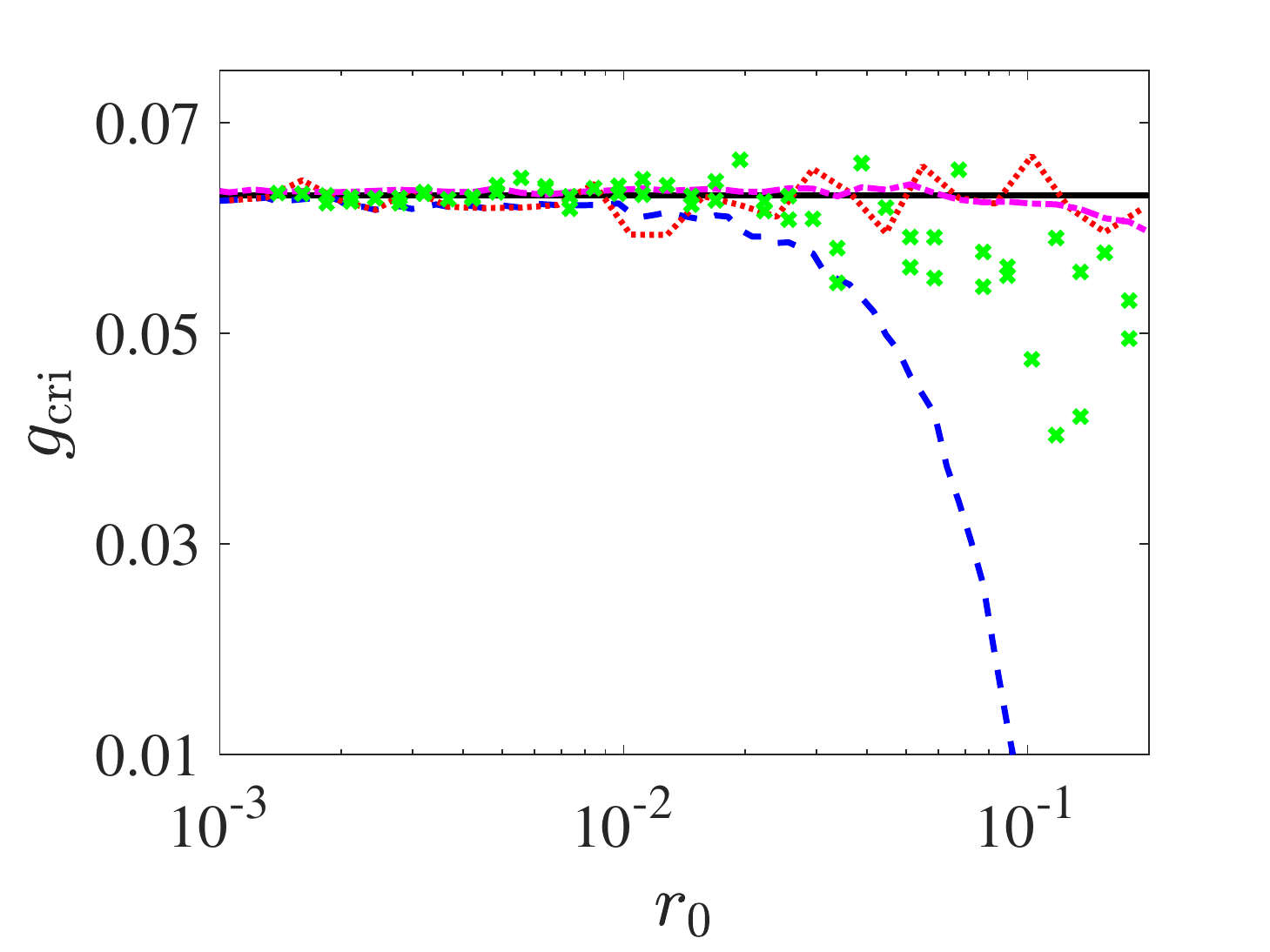}
\caption{
Critical points estimated via error mitigation (mean-field results). The black solid horizontal line, blue dashed line, red dotted line, cyan dash-dotted line, and cross symbols denote the results of values without error, with depolarizing error and linear extrapolation, with depolarizing error and quadratic extrapolation, with transverse damping error and with random Pauli error.
}
\end{figure}

\begin{figure*}[tbp]
\includegraphics[width=0.95\textwidth]{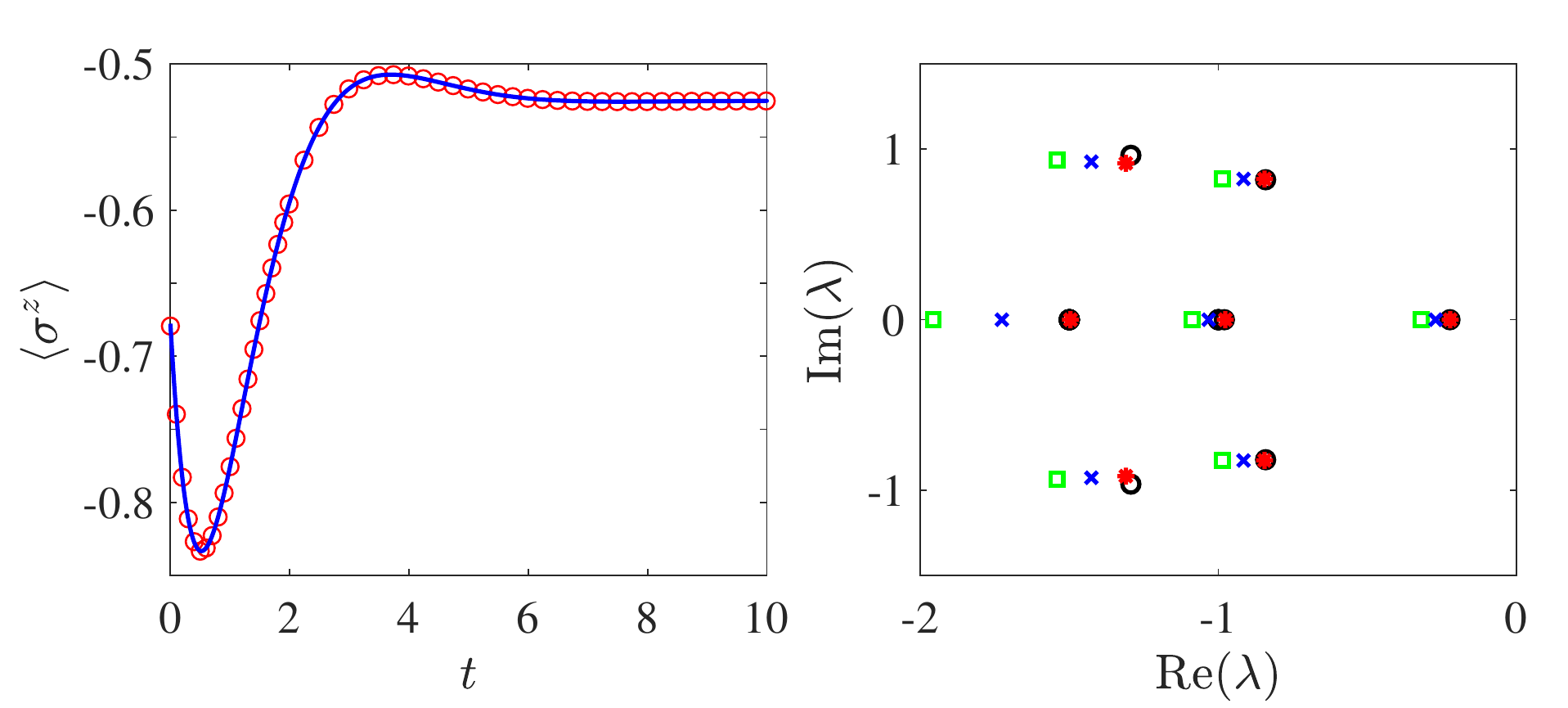}
\caption{The left panel shows an example of curve fitting by the matrix pencil method. The red circles mark the data of $\langle\sigma^z\rangle$ vs the evolution time $t$ extracted from the evolution of a randomly chosen initial density operator towards the steady state and the blue line is the analytical curve whose coefficients are extracted from the data by matrix pencil method. The right panel shows some eigenvalues of the Lindbladian. Blue cross and green squares are noisy values with $r=r_0=0.01$ and $r=2r_0=0.02$. Red asterisks are obtained from noisy values by zero-noise extrapolation, which matches the exact values denoted as black circles, which are obtained by directly diagonalizing the Lindbladian. The data are calculated at $g=0.1$ on a $3\times 3$ qubit lattice.}
\end{figure*}

\section{Spectroscopy} Finally, we show by numerical simulation that we can obtain spectroscopic information about the ideal Lindbladian ${\mathcal L}_0$ with a noisy quantum processor. Starting from a random initial density operator, we first calculate the nonequilibrium dynamics of expectation values of local physical observables, e.g., the local spin component $\mean{\hat\sigma^z}$, evolving towards the steady state $\hat\rho_{\rm eff}\left(r\right)$ on a noisy quantum process with the intrinsic and boosted noise rates $r_0$ and $cr_0$. These noisy nonequilibrium dynamics can be fitted with a multimodal decaying exponential function, i.e.,
\begin{eqnarray}
s(t)=\sum_{\alpha=1}^pA_\alpha e^{i\theta_\alpha}e^{\lambda_\alpha t},
\end{eqnarray}
where $A_\alpha\in{\mathbb R}$, $\theta_\alpha\in\left[0,2\pi\right]$ and $\lambda_\alpha\in{\mathbb C}$ are the parameters of the model, and we use the positive integer $p$ to control the complexity of the model. Note that ${\rm Re}\left(\lambda_\alpha\right)\leq 0$ for a physical Lindbladian. We use the matrix pencil method~\cite{sarkar1995using} to extract $\lambda_\alpha(r)$ from the nonequilibrium dynamics governed by ${\mathcal L}_{\rm eff}(r)$ with $r=r_0$ and $cr_0$. The error-mitigated estimation of the true eigenvalues of the target Lindbladian ${\mathcal L}_0$ can be obtained by linear extrapolation to the zero-noise point with $r=0$. By choosing different initial density operators and physical quantities in the nonequilibrium dynamics, we can extract most of eigenvalues of the target Lindbladian with small-magnitude real parts, while eigenvalues with large-magnitude real parts are difficult to extract due to their fast decay. As shown in Fig.~5(b), the extrapolated eigenvalues is well consistent with those of the true Lindbladian, i.e. $\lim_{r\rightarrow0}\lambda_{\alpha,\rm eff}(r)\simeq \lambda_\alpha$. 

\section{Conclusions} Starting with an initial state, we can simulate its time evolution via DQS in principle and extract any relevant physical quantity we want. The availability of only quantum computers constituting of noisy intermediate-scale quantum devices seems to prevent us from achieving this ambition. Due to the accumulation of errors, the simulation of long-time evolution is unreliable. The question is if we can find cases where physical quantities are not suffered from the breakdown of long-time evolution. In this paper, we show that the steady-state problem of open systems meet our demand. It is robut against error accumulation and we can extracted accurate values of observables via error mitigation. 

Our paper consists of two main parts. In the first part, we propose a perturbation theory, by which we prove that the deviation of the steady state depends only on the error in a single Troter step, regardless of the evolution time of DQS. This result has a simple physical interpretation. Error accumulation may corrupt the simulation of long-time evolution, but the steady state, as the goal of evolution, is not affected by it. The deviation of the noisy steady state from the exact one depends only on the difference between the noisy Lindbladian $\mathcal{L}_\mathrm{eff}$ and the exact one $\mathcal{L}_{0}$, which has nothing to do with the evolution time. Further, we can express the noisy steady state as a power series of the single-step error. Therefore, the exact steady state and any relevant observables can be obtained by error mitigation. In the second part, we take the dissipative XYZ model as an example to convince us of the theoretical results. When the strength of error is below a threshold, the perturbation theory works. By error mitigation and a technique based on scaling behavior, we can recover the exact values of magnetization, relaxation rate, and the location of the phase transition point. 


\section{Acknowledgement}
We acknowledge the support of the National Natural Science Foundation of China (Grant No. 11875050, 12088101, and 92065205) and NSAF (Grant No. U1930403).

\appendix

\section{Magnus expansion\label{app:me}}
According to Magnus expansion, the evolution operator can be expressed as 
\begin{eqnarray}
e^{\calL_{\rm eff} \tau} = \exp(\Omega_1 + \Omega_2 + \Omega_3 + \Omega_4 + \cdots),
\end{eqnarray}
where 
\begin{eqnarray}
\Omega_1 &=& \tau \sum_{j=1}^{N} \calG_j, \notag \\
\Omega_2 &=& \frac{\tau^2}{2} \sum_{j_1\geq j_2} M(j_1,j_2)^{-1} [\calG_{j_1}, \calG_{j_2}], \notag \\
\Omega_3 &=& \frac{\tau^3}{6} \sum_{j_1\geq j_2\geq j_3} M(j_1,j_2,j_3)^{-1} \notag \\
&&\times ([\calG_{j_1}, [\calG_{j_2}, \calG_{j_3}]] + [\calG_{j_3}, [\calG_{j_2}, \calG_{j_1}]]), \notag \\
\Omega_4 &=& \frac{\tau^4}{12} \sum_{j_1\geq j_2\geq j_3\geq j_4} M(j_1,j_2,j_3,j_4)^{-1} \notag \\
&&\times ([[[\calG_{j_1}, \calG_{j_2}], \calG_{j_3}], \calG_{j_4}] + [\calG_{j_1}, [[\calG_{j_2}, \calG_{j_3}], \calG_{j_4}]] \notag \\
&&+ [\calG_{j_1}, [\calG_{j_2}, [\calG_{j_3}, \calG_{j_4}]]] \notag \\
&&+ [\calG_{j_2}, [\calG_{j_3}, [\calG_{j_4}, \calG_{j_1}]]]).
\end{eqnarray}
Here, $M(j_1,j_2,\ldots) = \prod_{l=1}^N m_l!$, and $m_l$ is the occupation number of $l$, i.e.,~the number of $j$'s taking $j=l$. Then, the effective the effective Lindbladian reads $\calL_{\rm eff} = (\Omega_1 + \Omega_2 + \Omega_3 + \Omega_4 + \cdots)/\tau$. Expressing $\calL_{\rm eff}$ as a polynomial of $\tau$ and $r$, we have
\begin{eqnarray}
\calL_{\rm eff} = \calL_{\rm eff}^{(0)} + \calL_{\rm eff}^{(1)}+ \calL_{\rm eff}^{(2)} + \calL_{\rm eff}^{(3)} + \cdots \label{eq:leff}
\end{eqnarray}
where 
\begin{eqnarray}
\calL_{\rm eff}^{(0)} &=& \calL_0 = \sum_{j=1}^{N} \calG^{\rm id}_j, \notag \\
\calL_{\rm eff}^{(1)} &=& r \sum_{j=1}^{N} \calE_j + \frac{\tau}{2} \sum_{j_1\geq j_2} M(j_1,j_2)^{-1} [\calG^{\rm id}_{j_1}, \calG^{\rm id}_{j_2}], \notag \\
\calL_{\rm eff}^{(2)} &=& \frac{\tau r}{2} \sum_{j_1\geq j_2} M(j_1,j_2)^{-1} ([\calE_{j_1}, \calG^{\rm id}_{j_2}] + [\calG^{\rm id}_{j_1}, \calE_{j_2}]) \notag \\
&&+ \frac{\tau^2}{6} \sum_{j_1\geq j_2\geq j_3} M(j_1,j_2,j_3)^{-1} \notag \\
&&\times ([\calG^{\rm id}_{j_1}, [\calG^{\rm id}_{j_2}, \calG^{\rm id}_{j_3}]] + [\calG^{\rm id}_{j_3}, [\calG^{\rm id}_{j_2}, \calG^{\rm id}_{j_1}]]), \notag \\
\calL_{\rm eff}^{(3)} &=& \frac{\tau r^2}{2} \sum_{j_1\geq j_2} M(j_1,j_2)^{-1} [\calE_{j_1}, \calE_{j_2}] \notag \\
&&+ \frac{\tau^2 r}{6} \sum_{j_1\geq j_2\geq j_3} M(j_1,j_2,j_3)^{-1} \notag \\
&&\times ([\calE_{j_1}, [\calG^{\rm id}_{j_2}, \calG^{\rm id}_{j_3}]] + [\calG^{\rm id}_{j_3}, [\calG^{\rm id}_{j_2}, \calE_{j_1}]] \notag \\
&&+ [\calG^{\rm id}_{j_1}, [\calE_{j_2}, \calG^{\rm id}_{j_3}]] + [\calG^{\rm id}_{j_3}, [\calE_{j_2}, \calG^{\rm id}_{j_1}]] \notag \\
&&+ [\calG^{\rm id}_{j_1}, [\calG^{\rm id}_{j_2}, \calE_{j_3}]] + [\calE_{j_3}, [\calG^{\rm id}_{j_2}, \calG^{\rm id}_{j_1}]]) \notag \\
&&+ \frac{\tau^3}{12} \sum_{j_1\geq j_2\geq j_3\geq j_4} M(j_1,j_2,j_3,j_4)^{-1} \notag \\
&&\times ([[[\calG^{\rm id}_{j_1}, \calG^{\rm id}_{j_2}], \calG^{\rm id}_{j_3}], \calG^{\rm id}_{j_4}] + [\calG^{\rm id}_{j_1}, [[\calG^{\rm id}_{j_2}, \calG^{\rm id}_{j_3}], \calG^{\rm id}_{j_4}]] \notag \\
&&+ [\calG^{\rm id}_{j_1}, [\calG^{\rm id}_{j_2}, [\calG^{\rm id}_{j_3}, \calG^{\rm id}_{j_4}]]] \notag \\
&&+ [\calG^{\rm id}_{j_2}, [\calG^{\rm id}_{j_3}, [\calG^{\rm id}_{j_4}, \calG^{\rm id}_{j_1}]]]).
\end{eqnarray}

\section{Jordan normal form of Lindbladian and time evolution superoperator\label{app:jf}} For convenience, we express the Lindbladian in the Hilbert-Schmidt space\cite{Braggio2006full, Flindt2008counting, Emary2009counting, Watrous2018the}. We use $\HSket{A}$ to denote the $d^2$-dimensional column vector corresponding to each $d\times d$ matrix $A$, such that for two square matrices $A$ and $B$ the inner product $\HSbraket{A}{B} = \Tr(A^\dag B)$, where $\HSbra{A} = \HSket{A}^\dag$. For any orthonormal basis of the Hilbert space $\{ \ket{e_n} \}$, the set $\{ \ketbra{e_m}{e_n} \}$ is the orthonormal basis of the Hilbert-Schmidt space. With this basis, $\HSket{A} = \sum_{m,n} a_{m,n} \HSket{e_m,e_n}$ if $A = \sum_{m,n} a_{m,n} \ketbra{e_m}{e_n}$, where $\{ \HSket{e_m,e_n} \}$ is orthonormal. Then, the Lindbladian $\calL_0 = -i[H_0,\bullet] + \sum_i ( A_i \bullet A_i^\dag - \frac{1}{2} \{ A_i^\dag A_i,\bullet \} )$ can be expressed as a $d^2\times d^2$ matrix $\boldsymbol{L}_0 = -i(H_0 \otimes \openone_d - \openone_d \otimes H_0^{\rm T})  + \sum_i [ A_i \otimes A_i^* - \frac{1}{2} ( A_i^\dag A_i \otimes \openone_d + \openone_d \otimes A_i^{\rm T} A_i^* )$. Here, $\openone_d$ is the $d$-dimensional identity matrix. Using the Jordan normal form, we can write $\boldsymbol{L}_0$ in the block diagonal form  
\begin{eqnarray}
\boldsymbol{L}_0 &=& \sum_\alpha \lambda_\alpha \left(\sum_{m=1}^{d_\alpha} \HSketbra{R_{\alpha,m}}{L_{\alpha,m}} \right. \notag \\
&&\left. + \sum_{m=1}^{d_\alpha-1} \HSketbra{R_{\alpha,m}}{L_{\alpha,m+1}} \right),
\label{eq:JNF}
\end{eqnarray}
where each $\alpha$ denotes a Jordan block, $\lambda_\alpha$ is the corresponding eigenvalue, $d_\alpha$ is the dimension of the block, and $\sum_\alpha d_\alpha = d^2$. Here, $\{ \HSket{R_{\alpha,m}} \}$ and $\{ \HSbra{L_{\alpha,m}} \}$ are dimensionless vectors satisfying $\HSbraket{L_{\alpha,m}}{R_{\alpha',m'}} = \delta_{\alpha,\alpha'}\delta_{m,m'}$ and $\sum_{\alpha,m} \HSketbra{R_{\alpha,m}}{L_{\alpha,m}} = \openone_{d^2}$. 

For each Jordan block, the two vectors $\HSket{R_{\alpha,1}}$ and $\HSbra{L_{\alpha,d_\alpha}}$ are right and left eigenvectors of $\boldsymbol{L}_0$, i.e.~$\boldsymbol{L}_0 \HSket{R_{\alpha,1}} = \lambda_\alpha \HSket{R_{\alpha,1}}$ and $\HSbra{L_{\alpha,d_\alpha}} \boldsymbol{L}_0 = \HSbra{L_{\alpha,d_\alpha}} \lambda_\alpha$. All other vectors, i.e.~generalised eigenvectors, can be derived from the two using $\boldsymbol{L}_0 \HSket{R_{\alpha,m}} = \lambda_\alpha (\HSket{R_{\alpha,m}} + \HSket{R_{\alpha,m-1}})$ (where $m = 2,\ldots,d_\alpha$) and $\HSbra{L_{\alpha,m}} \boldsymbol{L}_0 = (\HSbra{L_{\alpha,m}} + \HSbra{L_{\alpha,m+1}}) \lambda_\alpha$ (where $m = 1,\ldots,d_\alpha-1$). 

In the standard Jordan normal form, off-diagonal elements are $1$. In Eq.~(\ref{eq:JNF}), off-diagonal elements are $\lambda_\alpha$. We choose this modified Jordan normal form such that eigenvectors $\{ \HSket{R_{\alpha,m}} \}$ and $\{ \HSbra{L_{\alpha,m}} \}$ are dimensionless. The modified Jordan normal form is possible because we always have $d_\alpha = 1$ when $\lambda_\alpha = 0$. 

Eigenvalues of a Lindbladian satisfy $\Re(\lambda_\alpha) \leq 0$. A Jordan block must be one-dimensional, i.e.~$d_\alpha = 1$, if $\Re(\lambda_\alpha) = 0$. Without loss the generality, we suppose that the first Jordan block corresponds to the unique steady state, i.e.,~$\lambda_1 = 0$ and $\HSket{R_{1,1}} = \HSket{\rho_0}$. Then, $\Re(\lambda_\alpha) < 0$ for all other Jordan blocks ($\alpha \neq 1$). Because $\Tr(\openone_d \calL_0(\bullet)) = 0$, i.e.,~$\HSbra{\openone_d} \boldsymbol{L}_0 = 0$, we have $\HSbraket{\openone_d}{R_{\alpha,m}} = 0$ for all $\alpha \neq 1$. We note that $\HSbra{L_{1,1}} = \HSbra{\openone_d}$. 

The evolution superoperator of $\calL_0$ is $\calV_0(t) = e^{\calL_0 t}$. Using the Jordan normal form, the corresponding matrix representation is 
\begin{eqnarray}
\boldsymbol{V}_0(t) &=& \sum_\alpha e^{\lambda_\alpha t} \sum_{m=0}^{d_\alpha-1} \frac{(\lambda_\alpha t)^m}{m!} \notag \\
&&\times \sum_{l=1}^{d_\alpha-m} \HSketbra{R_{\alpha,l}}{L_{\alpha,l+m}}.
\end{eqnarray}
Therefore, the state evolves to the steady state as $\norm{e^{\calL_0 t}(\hat\rho_{\rm in}) - \hat\rho_0} = {\mathcal O}\left(e^{-\Gamma t} {\rm poly}(t)\right)$, where $\Gamma = {\rm min}\left\{-{\rm Re}\left(\lambda_\alpha\right)|\alpha\neq 0\right\}$ determines the speed of the convergence, where $\lambda_\alpha$'s are the eigenvalues of ${\mathbf L}_0$.

\section{Perturbation theory\label{app:pt}}

In this section, we derive the perturbation theory of Lindbladian based on the modified Jordan normal form. 

\subsection{Steady state perturbation theory}

We express the steady state as $\hat \rho_{\rm eff} = \sum_{k=0}^\infty \hat\rho^{(k)}$, where $\hat\rho^{(k)}$ is the $k$-th-order correction to the steady state. Substituting this expression of $\hat\rho_{\rm eff}$ into $\calL_{\rm eff} \hat\rho_{\rm eff} = 0$ gives $(\calL_0 + \calL') (\sum_{k=0}^\infty \hat\rho^{(k)}) = 0$. Retain terms of the $k$-th order, we have $\calL_0(\hat\rho^{(k)}) + \calL'(\hat\rho^{(k-1)}) = 0$. Here, $\hat\rho^{(-1)} = 0$ and $\hat\rho^{(0)} = \hat\rho_0$. 

Now, we switch to the Hilbert-Schmidt space, and the equation of each order becomes 
\begin{eqnarray}
\boldsymbol{L}_0 \HSket{\rho^{(k)}} = - \boldsymbol{L}' \HSket{\rho^{(k-1)}},
\end{eqnarray}
where $\boldsymbol{L}_0$ is not invertible because of the zero eigenvalue $\lambda_1$. We need that $\hat\rho_{\rm eff}$ is normalized for each order of the correction, i.e.~$\Tr(\hat\rho^{(k)}) = \HSbraket{\openone_d}{\rho^{(k)}} = \HSbraket{L_{1,1}}{\rho^{(k)}} = 0$ if $k>0$. A generalized inverse of $\boldsymbol{L}_0$ is 
\begin{eqnarray}
\boldsymbol{L}_0^{-1} &=& \sum_{\alpha>1} \lambda_\alpha^{-1} \sum_{m=0}^{d_\alpha-1} (-1)^m \notag \\
&&\times \sum_{l=m+1}^{d_\alpha} \HSketbra{R_{\alpha,l-m}}{L_{\alpha,l}}.
\end{eqnarray}
Then, 
\begin{eqnarray}
\boldsymbol{L}_0\boldsymbol{L}_0^{-1} = \boldsymbol{L}_0^{-1}\boldsymbol{L}_0 &=& \sum_{\alpha>1} \sum_{m=1}^{d_\alpha} \HSketbra{R_{\alpha,m}}{L_{\alpha,m}} \notag \\
&=& \openone_d^2 - \HSketbra{R_{1,1}}{L_{1,1}}.
\end{eqnarray}
Therefore, 
\begin{eqnarray}
\HSket{\rho^{(k)}} &=& (\openone_d^2 - \HSketbra{R_{1,1}}{L_{1,1}}) \HSket{\rho^{(k)}} \notag \\
&=& \boldsymbol{L}_0^{-1}\boldsymbol{L}_0 \HSket{\rho^{(k)}} \notag \\
&=& - \boldsymbol{L}_0^{-1} \boldsymbol{L}' \HSket{\rho^{(k-1)}}.
\end{eqnarray}
We remark that this generalized inverse makes sure that $\hat\rho^{(k)}$ is traceless, i.e.~the steady state with a truncation at any order is normalized. 

\subsection{Non-degenerate perturbation theory}

We consider a non-degenerate eigenvalue $\lambda_\alpha$, i.e.,~$d_\alpha = 1$. Let $\lambda_\alpha^{(k)}$ and $\HSket{R_{\alpha,1}^{(k)}}$ be the $k$-th order corrections to the eigenvalue and right eigenvector, respectively. Then, 
\begin{eqnarray}
&&(\boldsymbol{L}_0+\boldsymbol{L}')(\HSket{R_{\alpha,1}^{(0)}} + \HSket{R_{\alpha,1}^{(1)}} + \HSket{R_{\alpha,1}^{(2)}} + \cdots) \notag \\
&=& (\lambda_\alpha^{(0)} + \lambda_\alpha^{(1)} + \lambda_\alpha^{(2)} + \cdots) \notag \\
&&\times (\HSket{R_{\alpha,1}^{(0)}} + \HSket{R_{\alpha,1}^{(1)}} + \HSket{R_{\alpha,1}^{(2)}} + \cdots),
\end{eqnarray}
where $\lambda_\alpha^{(0)} = \lambda_\alpha$ and $\HSket{R_{\alpha,1}^{(0)}} = \HSket{R_{\alpha,1}^{(k)}}$. 

Retain terms of the first order, we have 
\begin{eqnarray}
&&\boldsymbol{L}_0 \HSket{R_{\alpha,1}^{(1)}} + \boldsymbol{L}' \HSket{R_{\alpha,1}} \notag \\
&=& \lambda_\alpha \HSket{R_{\alpha,1}^{(1)}} + \lambda_\alpha^{(1)} \HSket{R_{\alpha,1}}.
\end{eqnarray}
Multiply this equation by $\HSbra{L_{\alpha,1}}$ from the left, we obtain the first-order correction to the eigenvalue 
\begin{eqnarray}
\lambda_\alpha^{(1)} = \HSbra{L_{\alpha,1}} \boldsymbol{L}' \HSket{R_{\alpha,1}}.
\end{eqnarray}
Multiple by $\HSbra{L_{\beta,m}}$ with $\beta\neq \alpha$, we have 
\begin{eqnarray}
&&\HSbraket{L_{\beta,m}}{R_{\alpha,1}^{(1)}} \notag \\
&=& \frac{\HSbra{L_{\beta,m}} \boldsymbol{L}' \HSket{R_{\alpha,1}} + \lambda_\beta \HSbraket{L_{\beta,m+1}}{R_{\alpha,1}^{(1)}}}{\lambda_\alpha - \lambda_\beta},
\end{eqnarray}
where $\HSbra{L_{\beta,d_\beta+1}} = 0$. Then, 
\begin{eqnarray}
\HSbraket{L_{\beta,m}}{R_{\alpha,1}^{(1)}} = \sum_{m'=m}^{d_\beta} \frac{\lambda_\beta^{m'-m} \HSbra{L_{\beta,m'}} \boldsymbol{L}' \HSket{R_{\alpha,1}}}{(\lambda_\alpha - \lambda_\beta)^{m'-m+1}},
\end{eqnarray}
Similar to the perturbation theory of Hamiltonian, we have $\HSbraket{L_{\alpha,1}}{R_{\alpha,1}^{(1)}} = 0$, such that the eigenvector after correction is still normalized, i.e.,~$\HSbraket{L_{{\rm eff},\alpha,1}}{R_{{\rm eff},\alpha,1}} = 1$, where $\HSket{R_{{\rm eff},\alpha,1}}$ and $\HSket{L_{{\rm eff},\alpha,1}}$ are right and left eigenvectors with the corrections. Then, we obtain the first order correction to the right eigenvector 
\begin{eqnarray}
&&\HSket{R_{\alpha,1}^{(1)}} \notag \\
&=& \sum_{\beta\neq \alpha,m}\sum_{m'=m}^{d_\beta} \frac{\lambda_\beta^{m'-m} \HSbra{L_{\beta,m'}} \boldsymbol{L}' \HSket{R_{\alpha,1}}}{(\lambda_\alpha - \lambda_\beta)^{m'-m+1}} \HSket{R_{\beta,m}}.
\end{eqnarray}
We can obtain the correction to the left eigenvector in a similar way. 

Retain terms of the second order, we have 
\begin{eqnarray}
&&\boldsymbol{L}_0 \HSket{R_{\alpha,1}^{(2)}} + \boldsymbol{L}' \HSket{R_{\alpha,1}^{(1)}} \notag \\
&=& \lambda_\alpha \HSket{R_{\alpha,1}^{(2)}} + \lambda_\alpha^{(1)} \HSket{R_{\alpha,1}^{(1)}} + \lambda_\alpha^{(2)} \HSket{R_{\alpha,1}}
\end{eqnarray}
Multiple this equation by $\HSbra{L_{\alpha,1}}$ from the left, we obtain the second order correction to the eigenvalue 
\begin{eqnarray}
\lambda_\alpha^{(2)} &=& \sum_{\beta\neq \alpha,m}\sum_{m'=m}^{d_\beta} \frac{\lambda_\beta^{m'-m} }{(\lambda_\alpha - \lambda_\beta)^{m'-m+1}} \notag \\
&&\times \HSbra{L_{\alpha,1}} \boldsymbol{L}' \HSket{R_{\beta,m}} \HSbra{L_{\beta,m'}} \boldsymbol{L}' \HSket{R_{\alpha,1}}.
\end{eqnarray}

\section{Circuits\label{app:c}}

\begin{figure}[tbp]
\centering
\includegraphics[width=1\linewidth]{\figpath 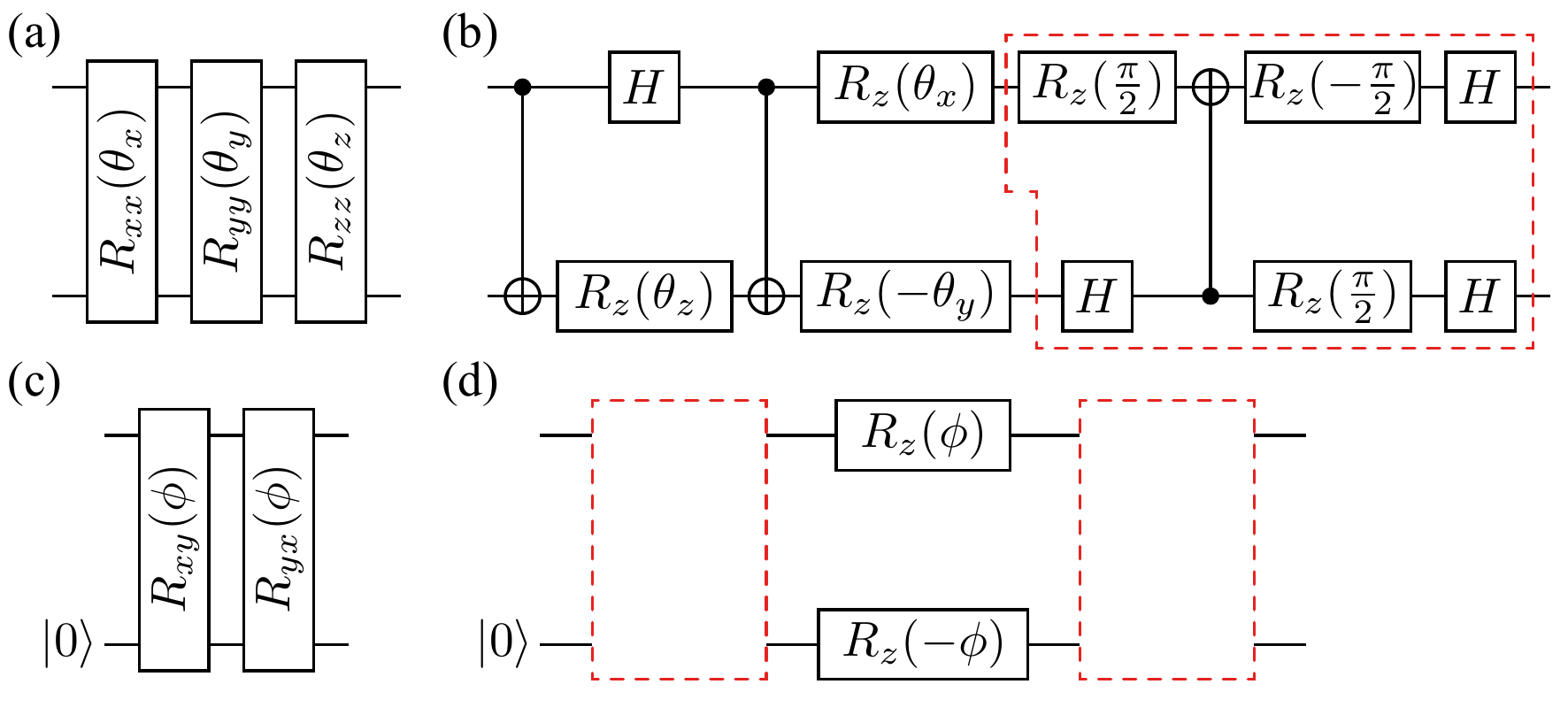}
\caption{
Circuits for implementing $e^{-i[H_{k,l},\bullet]\tau}$ and $e^{\calD_l\tau}$. (a) The circuit of $e^{-i[H_{k,l},\bullet]\tau}$ using small-angle gates. (b) The circuit of $e^{-i[H_{k,l},\bullet]\tau}$ using controlled-NOT gate. (c) The circuit of $e^{\calD_l\tau}$ using small-angle gates. The operation $e^{\calD_l\tau}$ is on the top qubit, and the ancillary qubit is initialized in the state $\ket{0}$. (d) The circuit of $e^{\calD_l\tau}$ using controlled-NOT gate. The dashed box represents the same circuit as in (b). 
}
\label{fig:circuit1}
\end{figure}

In this appendix we show the circuits for implementing the quantum gates for the dissipative XYZ model in Fig.~\ref{fig:circuit1}. Here and below, we omit the details of the qubit-ancilla permutation in $L\times 2L$ lattice which has been discussed in the main part and consider only the realization of two-qubit gate acting on qubit $k$ and $l$ and single-qubit acting on qubit $k$. We consider two setups. In the first setup, operations are realized using small-angle gates $R_{\alpha\alpha}(\theta) = \exp\left(-i\frac{\theta}{2}\sigma^{\alpha}\otimes\sigma^{\alpha}\right)$. We remark that $\theta_\alpha = 2J_\alpha\tau$ and $\cos\phi = e^{-\gamma \tau/2}$, and $\theta_\alpha,\phi \ll 1$ in the Trotterization algorithm. In the second setup, operations are realized using controlled-NOT gate and single-qubit gates. The single-qubit rotation gate reads $R_z(\theta) = \exp\left(-i\frac{\theta}{2}\sigma^z\right)$. 

\section{Error twirling and boosting\label{app:et}}

\begin{figure}[tbp]
\centering
\includegraphics[width=1\linewidth]{\figpath 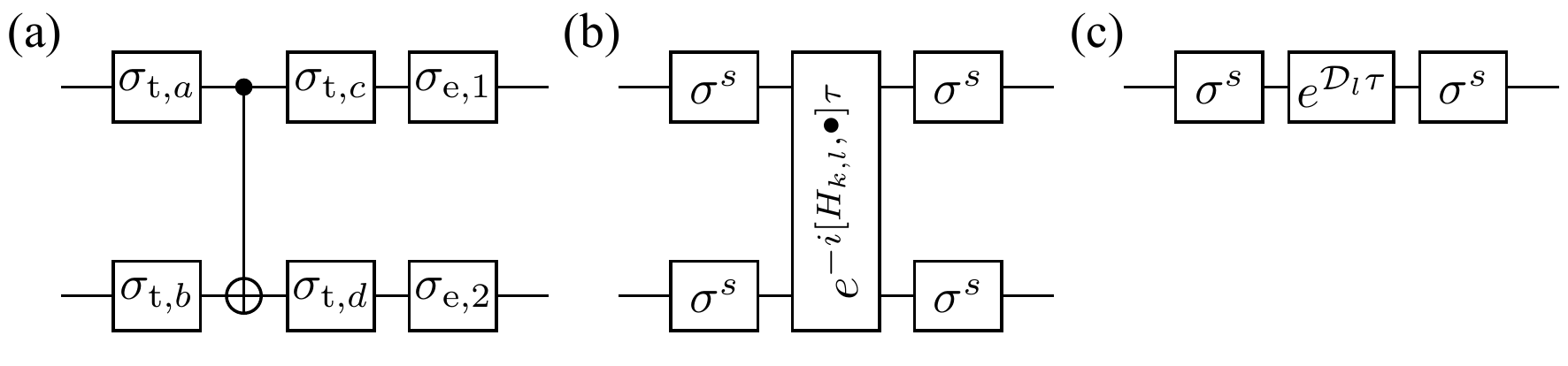}
\caption{
Pauli twirling and error boosting circuits. (a) Twirling and error boosting circuit for controlled-NOT gate. (b) Twirling circuit for $e^{-i[H_{k,l},\bullet]\tau}$. (c) Twirling circuit for $e^{\calD_l\tau}$. 
}
\label{fig:circuit2}
\end{figure}

The Pauli twirling of controlled-NOT gate uses four Pauli gates $\sigma_{{\rm t},q} = \sigma^I,\sigma^x,\sigma^y,\sigma^z$, where $q=a,b,c,d$, as shown in Fig.~\ref{fig:circuit2}(a). We randomly choose Pauli gates $\sigma_{{\rm t},a}$ and $\sigma_{{\rm t},c}$ according to the uniform distribution, and then we take $\sigma_{{\rm t},c}\otimes\sigma_{{\rm t},d} = \Lambda_x \sigma_{{\rm t},a}\otimes\sigma_{{\rm t},b} \Lambda_x$, where $\Lambda_x$ is the unitary operator of controlled-NOT gate. By applying the Pauli twirling, we can convert a general error model into the Pauli error model in the form $\calN [\Lambda_x]$, where $[\Lambda_x]$ denotes the completely-positive map of the ideal controlled-NOT gate, and $\calN = \sum_{\alpha,\beta = I,x,y,z} p_{\alpha,\beta} \left[ \sigma^{\alpha}\otimes\sigma^{\beta} \right]$ denotes Pauli errors. Here, $p_{I,I}$ is the fidelity of the gate, and $p_{\alpha,\beta}$ [$(\alpha,\beta)\neq (I,I)$] is the probability of the Pauli error $\sigma^{\alpha}\otimes\sigma^{\beta}$. To implement the error extrapolation, we need to measure error probabilities $p_{\alpha,\beta}$ using methods such as quantum process tomography. 

Suppose we want to effectively realize the error model $\calN' = \sum_{\alpha,\beta = I,x,y,z} p'_{\alpha,\beta} \left[ \sigma^{\alpha}\otimes\sigma^{\beta} \right]$, we can compute $\calN_{\rm e} = \calN^{-1} \calN'$. Note that the map $\calN$ is always invertible for high-fidelity gate. $\calN_{\rm e}$ is always a Pauli-error map in the form $\calN_{\rm e} = \sum_{\alpha,\beta = I,x,y,z} q_{\alpha,\beta} \left[ \sigma^{\alpha}\otimes\sigma^{\beta} \right]$. We remark that $q_{\alpha,\beta}$ is not always positive. To effectively realize the depolarizing map, we take $p'_{I,I} = 1-\epsilon$ and $p_{\alpha,\beta} = \epsilon/15$ for $(\alpha,\beta)\neq (I,I)$. We can always find a finite $\epsilon$ such that all $q_{\alpha,\beta}$ are positive. Intuitively, we can take $\epsilon = 15\max\{ p_{\alpha,\beta} \st (\alpha,\beta)\neq (I,I) \}$. Similarly, by taking $\calN'$ as the error model with increased error (e.g.~corresponding to $2r_{\rm raw}$), we can boost the error. Given $\calN_{\rm e}$, we can realize $\calN'$ by applying Pauli gates $\sigma_{{\rm e},1}\otimes\sigma_{{\rm e},2} = \sigma^{\alpha}\otimes\sigma^{\beta}$ with the probability $q_{\alpha,\beta}$, as shown in Fig.~\ref{fig:circuit2}(a). 

Pauli twirling circuit of the operation $e^{-i[H_{k,l},\bullet]\tau}$ is shown in Fig.~\ref{fig:circuit2}(b). This operation can be realised using either small-angle gates or controlled-NOT gate, and the twirling circuit works for both cases. To implement the twirling, we apply Pauli gates $\sigma^s = \sigma^I,\sigma^x,\sigma^y,\sigma^z$ on both qubits before and after the operation, according to the uniform distribution. Suppose $\calM_H = e^{\calL_H\tau}$ is the operation $e^{-i[H_{k,l},\bullet]\tau}$ with error after the twirling, we always have $\left[[\sigma^s\otimes\sigma^s], \calM_H \right] = \left[[\sigma^s\otimes\sigma^s], \calL_H \right] = 0$ for all $s$. 

It is similar for the operation $e^{\calD_l\tau}$. Pauli twirling circuit of the operation $e^{\calD_l\tau}$ is shown in Fig.~\ref{fig:circuit2}(c). This operation can be realized using either small-angle gates or controlled-NOT gate, and the twirling circuit works for both cases. To implement the twirling, we apply Pauli gates $\sigma^s = \sigma^I,\sigma^z$ before and after the operation, according to the uniform distribution. Suppose $\calM_D = e^{\calL_D\tau}$ is the operation $e^{\calD_l\tau}$ with error after the twirling, we always have $\left[[\sigma^z], \calM_D \right] = \left[[\sigma^z], \calL_D \right] = 0$. 

By using twirling circuits in Figs.~\ref{fig:circuit2}(b) and (c), we have either $\left[[\sigma^z\otimes\sigma^z] ,\calE_j\right] = 0$ or $\left[[\sigma^z] ,\calE_j\right] = 0$. Then, $\left[\calZ ,\calE_j\right] = 0$ for all $j$, i.e.~gate error after the twirling preserves the symmetry. 

For depolarizing error on controlled-NOT gate, because the map of two-qubit depolarizing error commute with single-qubit gates and two-qubit gates on the same two qubits, the error model of whole circuits in Figs.~\ref{fig:circuit1}(b) and (d) is depolarizing. In this case, further twirling operations using circuits in Figs.~\ref{fig:circuit2}(b) and (c) are unnecessary. 

\section{Noise models\label{app:em}}
In this section, we show how we deal with the errors $r\mathcal E$ in the quantum operation.
Besides a description of depolarizing error which is an equal superposition of Pauli errors, we also include a general case of random Pauli error and transverse damping error which has no ${\mathbb Z}_2$ symmetry. At last, we make some comments on the non-Markovian noise.

\subsection{Depolarizing error} 
As described in the main part, the real operation used in the universal quantum processer is $e^{\calG_m\tau}$, with $\calG_m = \calG_m^{\rm id} + r_m\calE_m$ and the generator of depolarizing error is 
\begin{eqnarray}
{\mathcal E}^{\rm d}_m = \left[\frac{\openone_{{\mathcal A}_m}}{2^{\left|{\mathcal A}_m\right|}}{\rm Tr}_{{\mathcal A}_m}\left(\bullet\right)-\bullet\right],
\end{eqnarray}
where we use a superscript below "d" to denote the depolarizing error, and "r" for random Pauli error, "t" for transverse damping error in the next subsection. In each Trotter step of quantum algorithm, we apply the real operation 
\begin{eqnarray}
e^{\calG_m^{\rm d}\tau} = e^{(\calG_{m}^{\rm id}+r_m\calE_m^{\rm d})\tau} = e^{r_m\calE_m^{\rm d}\tau}e^{\calG_m^{\rm id}\tau}
\end{eqnarray}
on the system. We choose $r_m=r$ for both single-qubit gates and two-qubits gates without loss of generality. 

In the mean-field calculation, the effective Lindbladian $\calL_{\rm eff}$ in the equations of motion
\begin{eqnarray}
\frac{d}{dt}\mean{\hat\sigma_i^\alpha}={\rm Tr}\left[\hat\sigma_i^\alpha{\mathcal L}_{\rm eff}\hat\rho\right],
\end{eqnarray}
is evaluated using the Magnus expansion. The methods are the same for the next two cases of errors.

\subsection{Random Rauli error}
In the Pauli twirling of controlled-NOT gate, we randomly choose Pauli gate. However, $\mathcal U_l(\tau)$ and $e^{\calD_k\tau}$ are in general not Clifford, thus not all the Pauli gates can be used in error twirling. As the dissipative XYZ model has the $\mathcal Z_2$ symmetry, i.e.,~$[\calZ, \calL_0] = 0$, where $\calZ = \prod_l [\sigma_l^z]$ , we can choose $\calE_j$ such that either $\left[[\sigma^z\otimes\sigma^z] ,\calE_j\right] = 0$ or $\left[[\sigma^z] ,\calE_j\right] = 0$. If $\calE_j$ satisfies $\left[[\sigma_l^z] ,\calE_k\right] = 0$, it must have the property that $\left[\calD_l ,\calE_j\right] = 0$. And if $\calE_j$ commutes with $\sigma_a^z\otimes\sigma_b^z$, it also commutes with the corresponding $\mathcal U_l(\tau)$. This constraints can be made looser by allowing the errors to commute with either $\sigma_a^x\otimes\sigma_b^x$ and $\sigma_a^y\otimes\sigma_b^y$. These facts motivate us consider the random Pauli error $\calE_j^{\rm r}$ to be sum of all these $\pm\calE_j$ satisfying theses constraints with probabilities $p_j$. The probabilities $p_j$ is supposed to be a uniform distribution.

\subsection{Transverse damping error} In the above two error models, there is always a phase transition from paramagnetic phase to ferromagnetic phase. This is not the case when the noise breaks the $Z_2$ symmetry. Besides the amplitude damping in $z$ direction which is crucial for the Dissipative XYZ model, we suppose there may be small-amplitude damping in $x$ and $y$ direction,
\begin{eqnarray}
\calD_l^x &=& \left( \sigma_l^{x-} \bullet \sigma_l^{x+} - \frac{1}{2} \left\{ \sigma_l^{x+} \sigma_l^{x-}, \bullet \right\} \right), \\
\calD_l^y &=& \left( \sigma_l^{y-} \bullet \sigma_l^{y+} - \frac{1}{2} \left\{ \sigma_l^{y+} \sigma_l^{y-}, \bullet \right\} \right),
\end{eqnarray}
where 
\begin{eqnarray}
\sigma^{x\pm} &=& \frac{1}{2}\left(\sigma^z \mp i\sigma^y \right),\\
\sigma^{y\pm} &=& \frac{1}{2}\left(\sigma^z \pm i\sigma^y \right).
\end{eqnarray}
We define transverse damping error $\calE_j^{\rm t}$ as equal superposition of $\calD_l^x$ and $\calD_l^y$.

To compare the effects of different noises, We choose the norm of the noise to be $\norm{\calE_j}=\sqrt{Tr\left(M^\dag M\right)}$, where $M$ is the matrix of the noise under the Pauli basis. In the numerical results of our calculation, the strength of noise for depolarizing error is choose to be the coefficient $r$ in $r\calE_j^{\rm d}$. When considering the other two cases of error, an extra factor $\norm{\calE_j^{\rm r/t}}/\norm{\calE_j^{\rm d}}$ is absorbed in $r$.

\subsection{Non-Markovian noise}

A way to generalize the formalism to non-Markovian noise is introducing the environment, i.e.~the initial state $\rho_{\rm i}^{\rm SE}$ is a state of the composite system formed by the system (qubits) and the environment. We assume that the time evolution of the composite system is Markovian. Then, the completely positive map of each step is 
\begin{eqnarray}
e^{\calG_N^{\rm SE} \tau} \cdots e^{\calG_2^{\rm SE} \tau} e^{\calG_1^{\rm SE} \tau} \equiv e^{\calL_{\rm eff}^{\rm SE} \tau},
\end{eqnarray}
where $\calG_j^{\rm SE}$ and $\calL_{\rm eff}^{\rm SE}$ are superoperators acting on the composite system. Similar to the case of Markovian noise, the generator of each gate can be expressed as $\calG_j^{\rm SE} = \calG^{\rm i}_j\otimes[\openone_{\rm E}] + r\calE_j^{\rm SE}$, where $[\openone_{\rm E}]$ is the identity map on the environment, and $\calE_j^{\rm SE}$ is the erroneous superoperators on both the system and environment.

\section{Supplementary numerical results\label{app:nr}}
\begin{figure}[tbp]
\includegraphics[width=1.0\linewidth]{\figpath 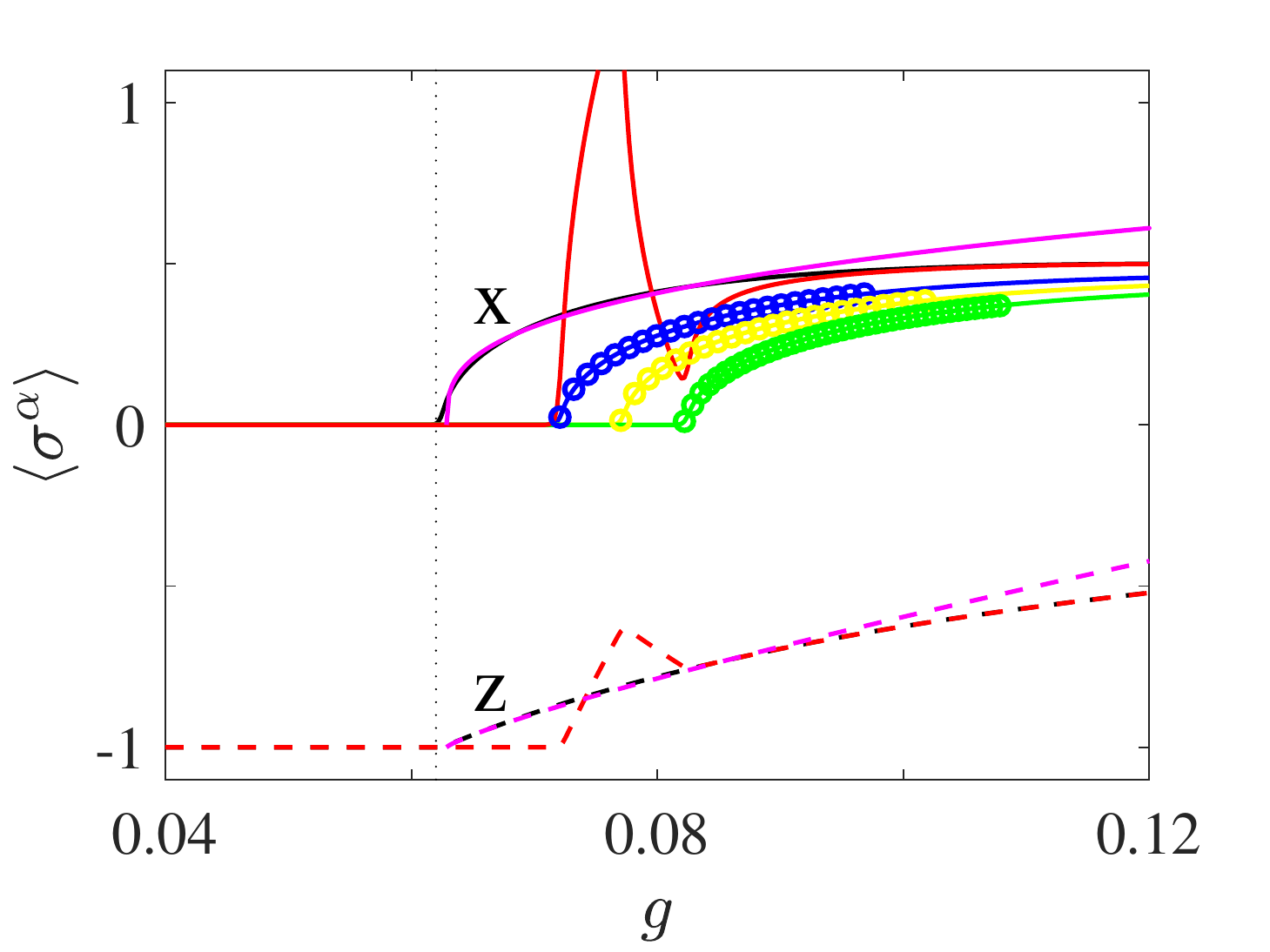}
\caption{
Zero-noise extrapolation and scaling extrapolation with quadratic ansatz. The strength is $r_0=0.01$ and $c=1$, $1.5$ and $2$ denoted by the blue, yellow, and green line. \label{fig:qua}
}
\end{figure}
The zero-noise extrapolation we do in the main part is of the first order. In concrete, we assume an observable of the noisy system depend linearly on the error strength $r$, i.e.,
\begin{eqnarray}
M(r) &= M_0 + b r,
\end{eqnarray}
where $M_0$ is the exact value of the observable, which we want. Then we choose two values of noise strength $r_0$ and $2r_0$, and by the numerical calculation, we can the values of $M(r_0)$ and $M(2r_0)$. It is easily shown that we get evaluate $M_0$ as
\begin{eqnarray}
M_0 &= 2M(r_0) - M(2r_0).
\end{eqnarray}
According to the perturbation theory, the steady state can be expressed as
\begin{equation}
\hat\rho_{\rm eff}=\sum_{n=0}^\infty(-1)^n\left({\mathcal L_0}^{-1}{\mathcal L'}\right)^n\hat\rho_0.
\end{equation}
Therefore, $M(r)$ is also a power series of the noise strength $r$, and the linear ansatz is the simplest approximation. In Fig.~\ref{fig:qua}, we show the zero extrapolation and scaling extrapolation with the quadratic ansatz 
\begin{eqnarray}
M(r) &= M_0 + b r + ar^2.
\end{eqnarray}

\begin{figure*}[htbp]
\centering
\includegraphics[width=0.95\textwidth]{\figpath 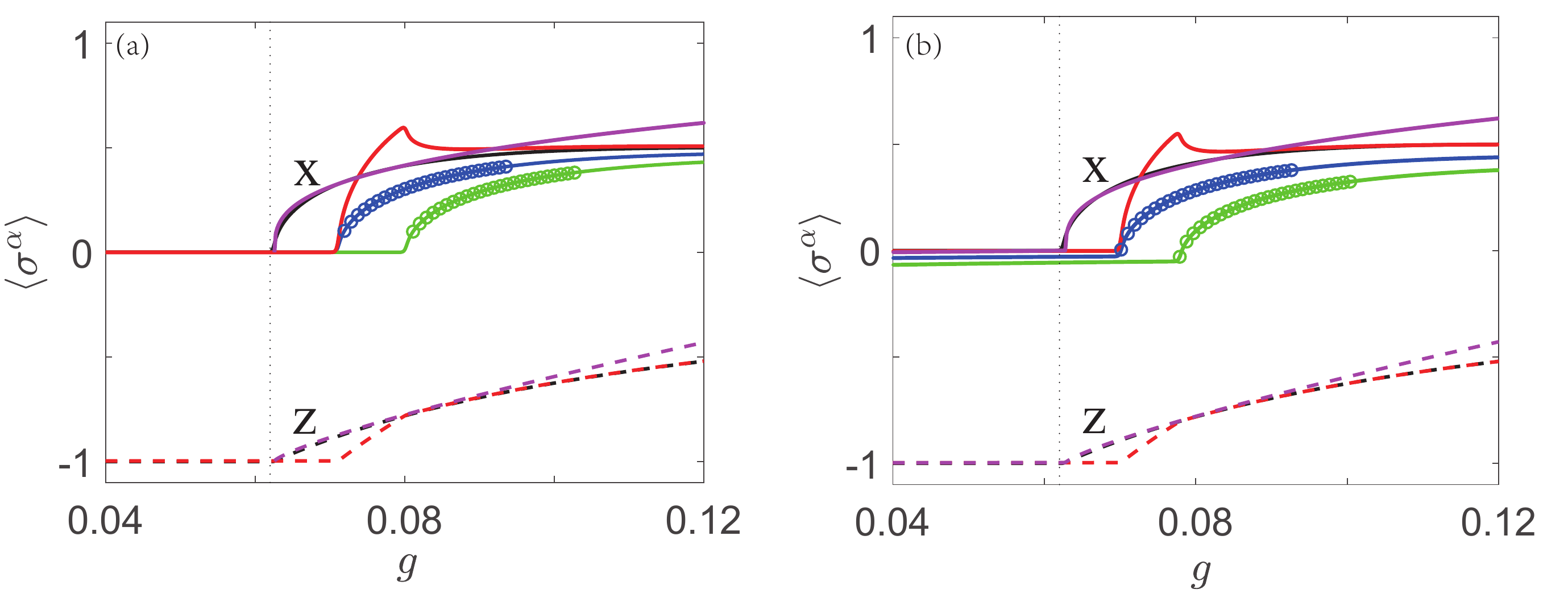}
\caption{
Zero-noise extrapolation of other noises (a) Random Pauli error. (b) Transverse damping error. In both plots, $r_0=0.01$ and $c=1,2$. \label{fig:noise}
}
\end{figure*}

In Fig.~\ref{fig:noise}, we show the zero-noise extrapolation and scaling extrapolation of random Pauli error and transverse damping error. It can be seen that the error without $\mathcal Z_2$ symmetry will destroy the phase transition. Therefore we modify Eq.~\eqref{eq:scal} as 
\begin{eqnarray}
m\left(g, r\right)\propto \left|g-g_{\rm cri}(r)\right|^{\beta(r)} + a(r) + b(r)r,
\end{eqnarray}
where $a(r)$ accounts for the deviation of "order parameter" from zero. There is a slight positive slope $b(r)$ which may not be directly seen from the plot.

\begin{figure}[tbp]
\centering
\includegraphics[width=1.0\linewidth]{\figpath 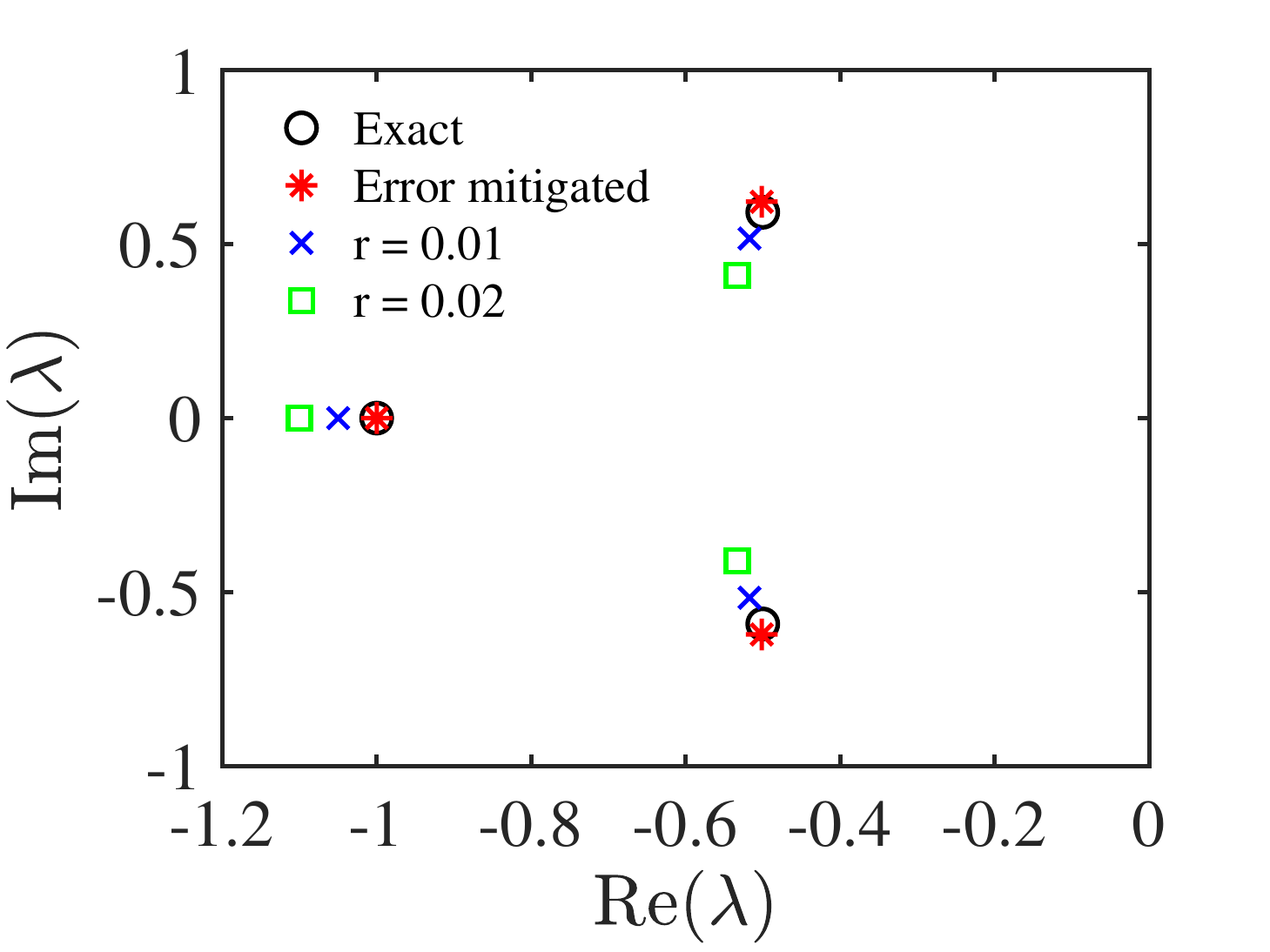}
\caption{
Eigenvalues obtained by the mean-field method.\label{fig:lmf}
}
\end{figure}

In Fig.~\ref{fig:lmf}, we show some lowest eigenvalues obtained by the mean-field method. Some of the eigenvalues differ much from that obtained on the $3\times 3$ qubit lattice. 

\end{document}